\documentclass[sigconf]{acmart}

\usepackage{CJK}
\usepackage[normalem]{ulem}
\usepackage{enumitem} 
\usepackage{amsmath}

\usepackage{balance}
\usepackage{float}
\usepackage{amssymb}
\usepackage{pifont} 
\usepackage[utf8]{inputenc}
\usepackage{tabularx} 
\usepackage{multirow} 
\usepackage{graphicx} 
\usepackage{xcolor}
\usepackage{colortbl}
\usepackage{arydshln}
\usepackage{subcaption}
\usepackage{listings}
\newcolumntype{L}[1]{>{\raggedright\let\newline\\\arraybackslash\hspace{0pt}}m{#1}}
\newcolumntype{C}[1]{>{\centering\let\newline\\\arraybackslash\hspace{0pt}}m{#1}}
\newcolumntype{R}[1]{>{\raggedleft\let\newline\\\arraybackslash\hspace{0pt}}m{#1}}

\definecolor{emphasis}{HTML}{84AEFF}
\definecolor{age}{HTML}{D79FF9}
\definecolor{gender}{HTML}{F274B8}
\definecolor{pitch}{HTML}{D8E637}
\definecolor{energy}{HTML}{FFA031}
\definecolor{speed}{HTML}{36E4E4}
\definecolor{topic}{HTML}{A3E584}
\definecolor{emotion}{HTML}{FFB2B2}

\AtBeginDocument{%
 }

\copyrightyear{2024}
\acmYear{2024}
\setcopyright{rightsretained}
\acmConference[MM '24]{Proceedings of the 32nd ACM International Conference on Multimedia}{October 28-November 1, 2024}{Melbourne, VIC, Australia}
\acmBooktitle{Proceedings of the 32nd ACM International Conference on Multimedia (MM '24), October 28-November 1, 2024, Melbourne, VIC, Australia}
\acmDOI{10.1145/3664647.3681674}
\acmISBN{979-8-4007-0686-8/24/10}

\makeatletter
\gdef\@copyrightpermission{
  \begin{minipage}{0.3\columnwidth}
   \href{https://creativecommons.org/licenses/by/4.0/}{\includegraphics[width=0.90\textwidth]{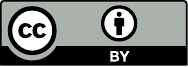}}
  \end{minipage}\hfill
  \begin{minipage}{0.7\columnwidth}
   \href{https://creativecommons.org/licenses/by/4.0/}{This work is licensed under a Creative Commons Attribution International 4.0 License.}
  \end{minipage}
  \vspace{5pt}
}
\makeatother

\begin{document}
\title[SpeechCraft: A Fine-grained Expressive Speech Dataset with Natural Language Description]{SpeechCraft: A Fine-grained Expressive Speech Dataset \\with Natural Language Description}


\author{Zeyu Jin}
\affiliation{%
 \institution{Department of Computer Science and
Technology\\Tsinghua University}
 \city{Beijing}
 \country{China}
}
\email{jinzeyu23@mails.tsinghua.edu.cn}

\author{Jia Jia}
\affiliation{%
 \institution{BNRist,Tsinghua University\\
 Key Laboratory of Pervasive Computing, Ministry of Education
 \city{Beijing}
  \country{China}
 }
}

\author{Qixin Wang}
\affiliation{%
 \institution{Department of Computer Science and
Technology\\Tsinghua University}
  \city{Beijing}
  \country{China}
}

\author{Kehan Li}
\affiliation{%
 \institution{Shenzhen International Graduate School\\
 Tsinghua University}
   \city{Shenzhen}
 \country{China}
}
 \author{Shuoyi Zhou}
\affiliation{%
  \institution{Shenzhen International Graduate School\\Tsinghua University}
    \city{Shenzhen}
 \country{China}
}
 \author{Songtao Zhou}
\affiliation{%
 \institution{Department of Computer Science and
Technology\\Tsinghua University}
  \city{Beijing}
 \country{China}
 }

\author{Xiaoyu Qin}
\orcid{0000-0002-9720-3220}
\authornote{~The corresponding authors: xyqin@tsinghua.edu.cn and zywu@sz.tsinghua.edu.cn.\\\textsuperscript{1} https://github.com/thuhcsi/SpeechCraft}
\affiliation{%
 \institution{Department of Computer Science and
Technology\\Tsinghua University}
   \city{Beijing}
 \country{China}
}

\author{Zhiyong Wu}
\authornotemark[1]
\affiliation{
\institution{
Shenzhen International Graduate School\\
Tsinghua University}
 \city{Shenzhen}
  \country{China}
}

\renewcommand{\shortauthors}{Zeyu Jin et al.}

\begin{abstract}

Speech-language multi-modal learning presents a significant challenge due to the fine nuanced information inherent in speech styles. Therefore, a large-scale dataset providing elaborate comprehension of speech style is urgently needed to facilitate insightful interplay between speech audio and natural language. However, constructing such datasets presents a major trade-off between large-scale data collection and high-quality annotation. To tackle this challenge, we propose an automatic speech annotation system for expressiveness interpretation that annotates in-the-wild speech clips with expressive and vivid human language descriptions. Initially, speech audios are processed by a series of expert classifiers and captioning models to capture diverse speech characteristics, followed by a fine-tuned LLaMA for customized annotation generation. Unlike previous tag/templet-based annotation frameworks with limited information and diversity, our system provides in-depth understandings of speech style through tailored natural language descriptions, thereby enabling accurate and voluminous data generation for large model training. With this system, we create \text{\textsc{SpeechCraft}$^1$}, a fine-grained bilingual expressive speech dataset. It is distinguished by highly descriptive natural language style prompts, containing approximately 2,000 hours of audio data and encompassing over two million speech clips. Extensive experiments demonstrate that the proposed dataset significantly boosts speech-language task performance in stylist speech synthesis and speech style understanding.

\end{abstract}



\maketitle

\section{Introduction}

\begin{table*}[t]
\newcounter{oldfootnote}
\setcounter{oldfootnote}{\value{footnote}}
\setcounter{footnote}{0}
\renewcommand{\thefootnote}{\fnsymbol{footnote}}
\centering
\caption{Examples of the style descriptions in different corpora. Each type of highlight/underline represents a unique speech attribute. 
In addition to the existing labels such as gender, pitch, volume, speed and emotional tone, descriptions in \textsc{SpeechCraft} involve properties of age, topic, emphasis, and transcript for the first time.}
\label{table_comparison_of_different_tts_dataset}
\resizebox{\textwidth}{!}{
\begin{tabular}{@{}cC{3cm}C{3cm}C{4cm}C{9cm}@{}}
\toprule
\textbf{FSNR0\footnotemark[3]~\cite{kim_expressive_2021}}&\textbf{ NLSpeech\footnotemark[4]~\cite{yang_instructtts_2023}} & \textbf{PromptSpeech~\cite{guo_prompttts_2023}} & \textbf{TextrolSpeech~\cite{ji_textrolspeech_2024}} & \textbf{\textsc{SpeechCraft} (ours)}\\
\midrule
Seem \colorbox{emotion}{sad} 
& The tone of the \colorbox{emotion}{shock} question revealed the \colorbox{emotion}{sad} feelings. 
& A \colorbox{emotion}{distressful} \colorbox{gender}{male} sound appeared in \colorbox{energy}{low volume} 
& A \colorbox{emotion}{heartbroken} \colorbox{gender}{woman}'s voice, almost a \colorbox{energy}{murmur}, is \colorbox{pitch}{high-pitched}. 
& Reflecting on a topic in the fields of \colorbox{topic}{Health and Fitness}, a \colorbox{emotion}{sad} \colorbox{age}{youth} with \colorbox{pitch}{low pitch} and \colorbox{energy}{normal volume} states, \uuline{"Well, you know, life is holistic, Dave."} \colorbox{gender}{She} speaks at \colorbox{speed}{a fast pace}, signifying her \colorbox{emotion}{sadness}. 
\\ \hline
In a \colorbox{speed}{hurry} 
& \colorbox{gender}{His} voice grew more \colorbox{emotion}{agitated}, and his tone revealed an \colorbox{speed}{urge and urgency}. 
& \colorbox{gender}{Men}, \colorbox{pitch}{low tone}, said \colorbox{energy}{loudly} and \colorbox{speed}{quickly} 
& Speaking at \colorbox{speed}{a fast pace}, the \colorbox{emotion}{pleasing} \colorbox{gender}{male} sustains a \colorbox{pitch}{regular pitch} and \colorbox{energy}{energy}. 
& Engaging in a conversation about \colorbox{topic}{portraits}, a \colorbox{emotion}{natural} \colorbox{age}{youth} \colorbox{gender}{female} with \colorbox{pitch}{high pitch} and \colorbox{energy}{normal volume} speaks \colorbox{speed}{swiftly}, describing:\uuline{"All his portraits seem to proclaim what a gentleman he is, and how he fascinates women!"}, intensifying the articulation of \colorbox{emphasis}{"fascinates"}.
\\
\bottomrule
\end{tabular}
}
\begin{flushleft}
    \footnotesize
    \quad\footnotemark[3] The FSNR0~\cite{kim_expressive_2021} style tags are the officially translated version from Korean.
    \footnotemark[4] The NLSpeech~\cite{yang_instructtts_2023} style prompts are the officially translated version from Chinese.
\end{flushleft}
\setcounter{footnote}{\value{oldfootnote}}
\renewcommand{\thefootnote}{\arabic{footnote}}
\end{table*}

The success of \textit{multi-modal learning}~\cite{xu_multimodal_2023} has boosted a swift resurgence in the development of speech-language models over recent years, 
encompassing improvements in speech synthesis and automated audio captioning. 
Large-scale text-to-speech (TTS) models (e.g., VALL-E~\cite{wang_neural_2023}, Natural Speech~2~\cite{shen_naturalspeech_2023}) and audio-to-text (ATT) models (e.g., SALMONN~\cite{tang_salmonn_2023} and  Qwen~\cite{bai_qwen_2023}) not only excel in traditional tasks through a \textit{zero-shot fashion} but also exhibit emergent behaviors. 
The utilization of vast quantities of high-quality labeled data in training plays a crucial role in facilitating these advancements~\cite{zhou_machine_2017}.
However, existing research mainly focused on fundamental audio characteristics, e.g., producing intelligible speech ~\cite{ren_fastspeech_2020, kong_vits2_2023, shen_natural_2018} and classifying broad sound event~\cite{kumar_audio_2016, chan_comprehensive_2020}. 
In contrast, nuanced audio interpretation that delves into the finer details of speech, particularly the speaking style, remains less explored. 

The \textit{style} of speech encompasses not only prosody but also the speaker's identity, emotional undertones, contextual cues, and scenes related to the topic within an audio clip~\cite{vyas_audiobox_2023}.
Current TTS systems lack the necessary flexibility for precise and disentangled control over speech style.
Meanwhile, current audio captioning systems struggle to capture the finer nuances besides rough detection of `\textit{a man is talking while a dog is barking}'. 

Nevertheless, few open-source datasets provide extensive details on vocal characteristics and nuanced descriptions~\cite{ji_textrolspeech_2024}. 
The limited scale of existing fine-grained datasets significantly impedes speech-language style research.
Consequently, there is an urgent demand for data that characterizes rich and detailed vocal information with natural language, 
both for expressive \textit{speech language understanding} (SLU) and \textit{controllable speech synthesis}.




It is widely acknowledged that human annotation datasets are typically costly, time-consuming, and limited in scope.
To tackle the trade-off between large-scale data collection and high-quality annotation, we develop an \textit{automatic speech annotation system for expressiveness understanding}, 
which conducts an exhaustive analysis of unlabeled audio across various dimensions 
and generates customized natural human language descriptions.
The system incorporates expert classifiers and sophisticated captioning models to determine multiple speech attributes. 
Notably, for the first time, 
we not only take into consideration the basic speaking properties such as gender, emotion and pitch, but also pay attention to the detailed prosodic characteristics including word emphasis and topic information. 
Leveraging the extraordinary abilities of LLMs in language comprehension and generation, we employed a fine-tuned LLaMA~2~\cite{touvron_llama_2023}  to integrate attributes into comprehensive and stylistic descriptions. 
The produced descriptions are tailored for each audio piece, as opposed to previous works that employ predefined templates to fill blanks with properties. 
The customization significantly enhances the diversity and nuances of the descriptions, aligning them with the unique characteristics of audio clips.

The proposed annotation system can encompass the most fine-grained attributes 
and the most diverse natural language descriptions available. 
Considering English and Chinese are the two most widely used languages globally, 
we applied the annotation system to
four popular bilingual speech datasets, including AISHELL-3~\cite{shi_aishell-3_2020}, Zhvoice\footnote{https://github.com/fighting41love/zhvoice}, LibriTTS-R~\cite{koizumi_libritts-r_2023}, GigaSpeech-m~\cite{chen_gigaspeech_2021}. 
This effort resulted in the creation of the largest open-source expressive speech dataset, named \textsc{SpeechCraft}, 
which comprises over 2,000 hours of audio data and more than two million speech clips. Examples as shown in Tab. \ref{table_comparison_of_different_tts_dataset}.
Experiments in speech-related tasks show that \textsc{SpeechCraft} dataset significantly contributes to the advancement of speech-language multi-modal learning in both TTS and speech-to-text (SST) domains. 
It enhances the performance of expressive speech synthesis, enables precise control over speech emphasis through natural language, and equips automated captioning systems with a broader context understanding capability to describe detailed speaking styles beyond mere speech event detection.


In summary, our contributions are threefold:
\begin{itemize}[noitemsep, topsep=0pt, partopsep=0pt,leftmargin=*]
\item We proposed an automatic speech annotation system that employs all-encompassing speech interpretation methods and a fine-tuned language model to cultivate highly descriptive comprehension of speech expressiveness.
\item We proposed an open-source, large-scale bilingual dataset available for advanced speech-language learning with fine-grained and expressive descriptions of speech, named \textsc{SpeechCraft}.
\item Leveraging the vast potential of \textsc{SpeechCraft}, we accomplished controllable speech synthesis with precise emphasis control, and automated captioning with detailed descriptions of acoustic properties and speaker identity for the first time.

\end{itemize}


\section{Related~Works}
\label{sect_relatedworks}

\begin{table*}[t]
\centering
\caption{Comparison between stylistic speech datasets. The proposed \textsc{SpeechCraft} dataset possesses larger scale and finer-grained properties.} 
\label{table_comparison_of_different_dataset}
\renewcommand{\thefootnote}{\fnsymbol{footnote}}
\resizebox{\textwidth}{!}{
\begin{tabular}{@{}cccccp{-3mm}cccc@{}}
\toprule
\multirow{2}{2cm}{\centering\textbf{Dataset}}& \multicolumn{3}{c}{\textbf{Size of Dataset}} & \multicolumn{6}{c}{\centering\textbf{Property of Dataset}} \\
\cline{2-5} \cline{7-10}
 & \textbf{\#Duration}& \textbf{\#Clips} & \textbf{\#Speakers} & \textbf{\#Labels} & & \textbf{Language} & \textbf{Audio Source} & \textbf{Description Form} & \textbf{Open Source} \\
\midrule
FSNR0~\cite{kim_expressive_2021} & 26h & 19k & 1 & 1 && KO & Internal dataset & Style tag & \text{\ding{51}}  \\
NLSpeech~\cite{yang_instructtts_2023} & 44h&- & 7 & 2 && ZH & Internal dataset & Human annotation & \text{\ding{55}}  \\
PromptSpeech~\cite{guo_prompttts_2023} & -&28k & - & 
5 && EN & AUD, TTS synthesized & Human annotation + LM rewrite & {\color{white}\footnotemark[1]}~\text{\ding{51}}~\footnotemark[6] \\
PromptTTS~2~\cite{leng_prompttts_2023} & -&20k & - & 
4&& EN & AUD & LLM template & \text{\ding{55}}  \\
TextrolSpeech~\cite{ji_textrolspeech_2024} & 330h & 236k & 1,324 & 
5 && EN & AUD, Emotional dataset & LLM template & \text{\ding{51}} \\
Audiobox~\cite{vyas_audiobox_2023} & >500h&- & - & 
8 && EN & Internal dataset & Human annotation + LLM rewrite & \text{\ding{55}}  \\
\multicolumn{1}{c}{\cellcolor{gray!25}\textsc{SpeechCraft} (ours)} & \cellcolor{gray!25}2,391h&\cellcolor{gray!25}2,250k & \cellcolor{gray!25}>3,200 & 
\cellcolor{gray!25}8 &\cellcolor{gray!25}& \cellcolor{gray!25}EN + ZH &\cellcolor{gray!25} AUD, YOU, POD, Smart Agent & \cellcolor{gray!25}LLM customization  for each piece & \multicolumn{1}{c}{{\color{gray!25}\footnotemark[1] }\cellcolor{gray!25}\text{\ding{51}}} \\
\bottomrule
\end{tabular}
}
\begin{flushleft}
    \footnotesize
    \quad\footnotemark[6] Approximately 85\% of the corpus in PromptSpeech~\cite{guo_prompttts_2023} was not released.
\end{flushleft}
\renewcommand{\thefootnote}{\arabic{footnote}}
\end{table*}

\textit{Content} and \textit{style} are two aspects of vital importance in speech audio formation.
Numerous speech corpora have been released, such as LibriVox-based corpus~\cite{kearns_librivox_2014}; 
however, datasets annotated with speech styles are limited. 
Besides, existing datasets typically describe the style of speech using tags or templates. 
Such annotations may be insufficient in diversity and richness for training large-scale speech-related models.

In this section, we review existing stylistic speech datasets and related works on audio captioning, which is highly relevant to speech-style captioning as discussed in Sec.~\ref{sct:automated_speech_style_captioning}.


\subsection{Tag-based Speech Datasets}

Speech style control starts from the usage of speaker ID tags in traditional speech language datasets ~\cite{ardila2019common, kearns_librivox_2014}, which intrigued voice cloning and one-shot TTS. Subsequently, emotional speech datasets and multi-modal datasets, such as TESS~\cite{dupuis_toronto_2010}, SAVEE~\cite{vlasenko_combining_2007}, IEMOCAP~\cite{busso_iemocap_2008}, MEAD~\cite{wang_mead_2020}, etc., flourish the field of \textit{emotional speech synthesis} and \textit{speech emotion recognition} (SER). However, emotion is usually defined as a classification task with different categories by datasets.
The lack of an acknowledged definition in emotion classification indicates that classifying emotion in single-word categories is not enough to represent emotion nuances.
Moreover, the existing emotional data used for speech synthesis are mostly recorded in an acting style, which are usually far from daily speech.

\subsection{Natural Language Stylistic Datasets}
\label{subsec:Natural-Language-based Stylistic TTS Dataset}


Speech datasets with natural language style prompts extend the formation of emotional tags to better capture the emotion in speech. 
As the pioneer in text prompt speech synthesis, InstructTTS~\cite{yang_instructtts_2023} recruited human annotators to describe speech emotion from three levels.
Researchers then focused on increasing the variety of style factors and the diversity in style prompt descriptions to elaborate speech style. 
As human annotation is always costly and size-limited, LLMs drive rapid development.
PromptTTS~\cite{leng_prompttts_2023} employed five style factors and adopted SimBERT~\cite{su_simbert_2020} to generate more style prompt templates with similar semantics.
PromptTTS 2~\cite{leng_prompttts_2023} further promoted a prompt generation pipeline with an SLU part and a LLM part to compose high-quality text prompts. It developed diversity in vocabulary format and reused ChatGPT templates for utterances with the same labels.
MM-TTS~\cite{guan_mm-tts_2024} and TextrolSpeech~\cite{ji_textrolspeech_2024} adopted ChatGPT in generating various prompt templates. 
Notably, TextrolSpeech is so far the largest open-source stylistic speech corpus. 
It collected and curated a series of existing emotion datasets, and involved traditional TTS datasets as neutral emotion utterances. 
Comparison between stylistic speech datasets is shown in Tab. \ref{table_comparison_of_different_dataset}. 

However, the concrete categories of each style property limit the summation of total style permutation, since the templates do not provide additional speech style information. 
Taking TextrolSpeech as an example, descriptions are derived from five style factors including gender, pitch, speaking speed, volume, and emotion. The emotion factor has eight options while other factors have two to three options each, a total of 432 combinations. Similarly, PromptTTS~2 has 54 combinations and MM-TTS provided 48 combinations. 
As a result, two different audio utterances with the same labels and templates would become totally the same in after the prompt programming.
The key barrier is that the descriptions are created based on LLM's capability given barely the text attribute labels instead of real audio samples, thus LLM could not provide unique annotations for each speech segment.

Another highly descriptive style dataset is Audiobox~\cite{vyas_audiobox_2023}.
It leveraged all kinds of miscellaneous details with human annotation and a quality assurance rating system to gather better alignment towards human hearing perception. 
Yet, the data was unavailable and the method is by no means universally applicable.
Consequently,
there is still an urgent need to develop an automatic method that transcends the template-based description generation approach, 
achieving tailored style descriptions for individual audio pieces.


\begin{figure*}[h]
  \vspace{-0.25cm}
  \includegraphics[width=0.9\linewidth]{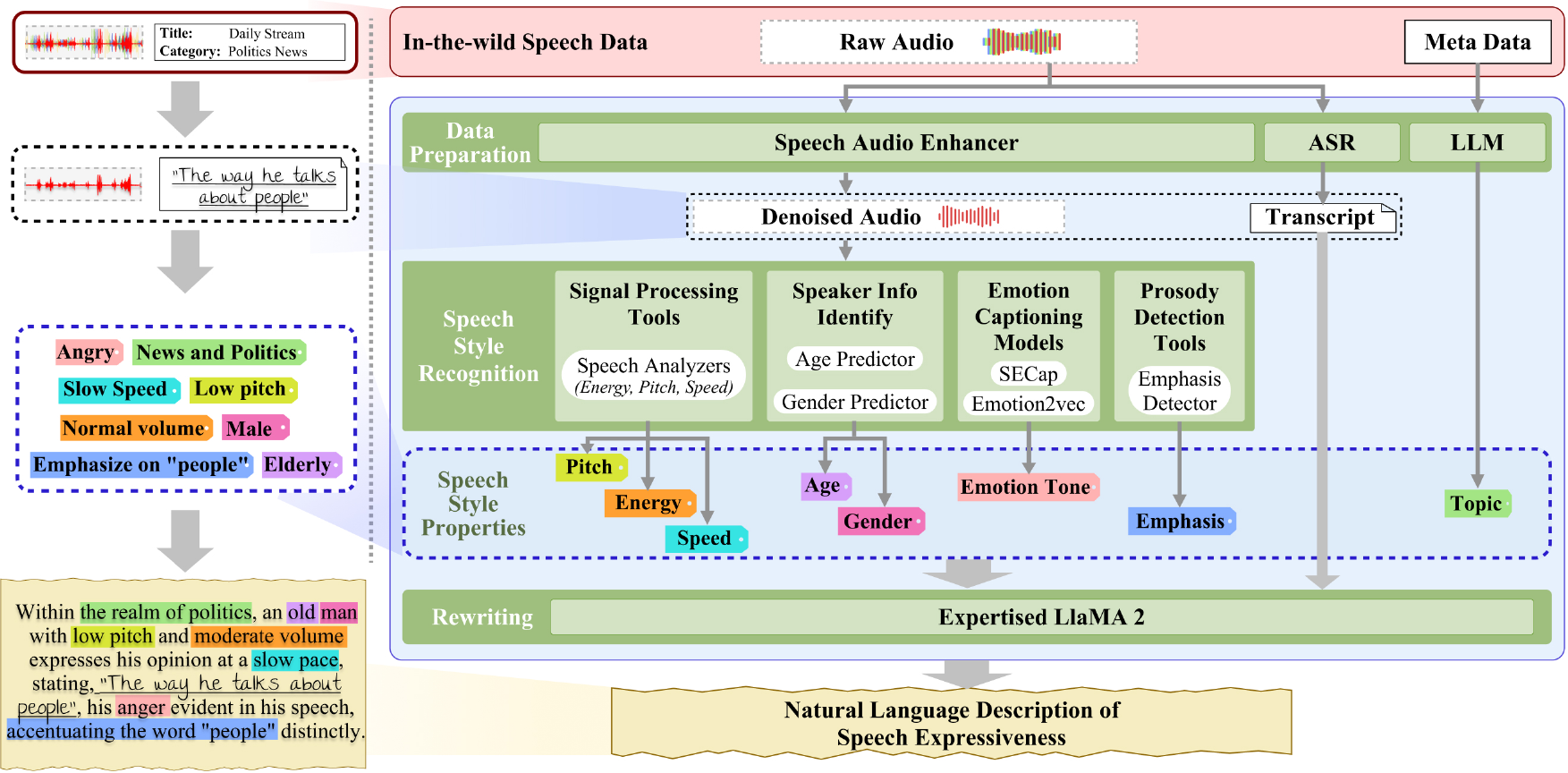}
  \vspace{-0.4cm}
  \caption{System framework of the automatic speech annotation system.}
  \label{fig:framework}
  \vspace{-0.25cm}
\end{figure*}

\subsection{Auto-captioning from Audio to Speech}

Audio or speech captioning refers to the task of generating descriptive text that describes the sounds, speech, and other contextual information of audio clips.
It can bridge the gap between auditory content and natural language, paving the way for deeper semantic connections in a variety of advanced audio related multimodal tasks~\cite{sun2024video, wang2024dancecamera3d, jin2023holosinger}.

Large-scale audio captioning datasets~\cite{kim_audiocaps_2019, drossos_clotho_2020, wu_large-scale_2023, mei_wavcaps_2023} enable the development of automated audio captioning (AAC) models to understand complex auditory scenes. 
AudioClip~\cite{guzhov_audioclip_2022} and CLAP~\cite{radford_learning_2021} employed contrastive learning paradigm to learn acoustic concepts from natural language supervision. The fellow AAC models surpassed traditional downstream audio tasks~\cite{koepke_audio_2023, chan_comprehensive_2020, kim_audiocaps_2019}.

As to speech captioning, 
AAC with natural language supervision casts light on captioning with speech emotion description.
Despite the lack of large-scale speech description data, SECap~\cite{xu_secap_2024} proposed a speech emotion captioning framework based on a small-scale human-annotated internal dataset.
It further employed the Q-former strategy to better disentangle the emotion-related speech information from the general semantic features.
We made a good application of SECap in our automatic speech annotation system to capture unlabeled speech data with detailed and unique descriptions in emotional tone.
It provides us with partial stylistic information which enables LLMs to directly interpret from authentic audio features.
Regrettably, to the best of our knowledge, current speech interpretation models fall short of addressing any stylistic dimensions beyond emotion in fine-grained natural language captions. 
Furthermore, there is an absence of a speech captioning model capable of describing the full spectrum of speech style within a complete sentence.







\label{sect_expressive_tts}


\section{Descriptive Speech Interpretation}
\label{sec:pagestyle}

\subsection{Overview of the Annotation System}

We propose an automatic speech annotation system for expressiveness interpretation.
It contains a three-stage data processing pipeline: data preparation for multiple sourced raw speech segments, property extraction with an all-aspect SLU framework, and customized rewriting with a fine-tuned LLM.
The system can equip in-the-wild speech with detailed and diverse descriptions.

\subsection{Data Preparation}

Multiple sourced raw audio varied in data quality and the form of original metadata. 
To reform the data with enhanced format, we first conduct data preprocessing to improve data quality, prepare data transcription, and establish standard metadata (metadata
example in supplementary materials and GitHub repository). 
All the speech segments, if not sourced from professional audiobooks, are fed into the speech enhancement system for audio quality improvement.
Parallel to audio enhancement, the content of speech audio is transcribed using Whisper Large-v3~\cite{radford_robust_2022} if not provided.
Other sundry items available in the original audio information such as title, raw descriptions from the data uploader, and video category tags from the website, are mutually transferred by a language model to summarize the topic of the speech utterance.

\subsection{Speech Style Recognition}
\label{sec:Speech Style Recognition}
The workflow of the speech style recognition is described as Fig.~\ref{fig:framework}. The speech is processed through various audio feature extraction models for characterizing speech in terms of its style properties. 
The output labels consist of pitch, energy, speed, age, gender, emotion description, and word emphasis, as illustrated in Fig.~\ref{fig:framework}.\\


\noindent\textbf{Signal Processing Tools.}
We utilize traditional signal processing tools to analyze audio signals and further predict acoustic properties such as pitch, energy, and speed. 
Speed and energy labels are categorized three-fold.
As to the pitch label, following the precedent standard of Audiobox~\cite{vyas_audiobox_2023}, we obey the common sense that female tends to have a higher pitch than male, setting categories of pitch by gender. 


\noindent\textbf{Speaker Information Identification.}
With the rapid development of large-scale audio representative learning methods, the high-level speaking features are well-detected by the hidden layers of powerful audio encoders. 
The pre-trained large-scale audio foundation models outperformed all kinds of downstream tasks compared to the respective task-specific models, including speaker features recognition and classification.
Wav2vec~2.0 ~\cite{baevski_wav2vec_2020}, attached with some additional linear layers, are fine-tuned for identifying speaker information like gender and age. 


\noindent\textbf{Emotion Caption.}
Emotional tone is the fundamental key to style in speech. 
It is important to preserve the nuances of emotional tone in the original audio to the fullest extent possible.
The audio utterances from the audiobooks are primarily storytelling narration with minimal emotion. 
In line with TextrolSpeech, we set their emotional tones to be `neutral'.
In the case of other emotion caption options, we take advantage of the most advanced emotion recognition technologies according to languages.
As to English audio data, 
we adopt a pre-trained speech emotion representation model, Emotion2vec~\cite{ma_emotion2vec_2023}. It provides convincing nine-class emotion recognition results and achieves SOTA on SER tasks.
For Chinese audio data, 
we adopt the speech emotion captioning model SECap~\cite{xu_secap_2024}. 
It captures the nuances of emotional cues, including intensity and fluctuations, through short sentences rather than predefined single words.
The customization of emotional tune makes the caption of each audio piece unique from others.
The content serves as the textual basis for the final speech expressiveness description.

\noindent\textbf{Word Emphasis Detection.}
\label{sec:Word Emphasis Detection}
Word emphasis plays a crucial role in speech expressiveness, conveying particular attitudes beyond the mere lexical content of what is spoken. 
Emphasis typically manifests as the strategic accentuation of certain words within a sentence. 
Therefore, the minimum unit for emphasis detection is set to be words in Chinese and English. 
Inspired by the lexical stress detection\footnote{https://github.com/LexicalStressDetection/lexical-stress-detection} in isolated English characters, we consider both the spectral and non-spectral features to hierarchically model the acoustic information. 
We model the spectral features with a residual convolutional neural network and non-spectral features with a deep neural network respectively. 
Details of the model implementation can be found in supplementary materials.
The emphasis of sentence is the word predicted with top~1 probability.

\subsection{Rewriting via LLMs }

Capitalizing on the exceptional annotation capabilities of LLMs, we employed an expertise LLaMA~2\footnote{Due to the original LLaMA’s limited proficiency in Chinese, we adopted an alternate version of LLaMA (Baichuan2-7B-Base~\cite{yang_baichuan_2023}) trained on bilingual corpus.} to transfer the group of attribute contents to natural language description for speech expressiveness. 
It is worth noting that we do not provide any structured formats for the description in advance to fill in the blanks as PromptTTS~2 does, but put emphasis on the richness of vocabulary and the accuracy in conveying the meaning of labels.
To regularize LLaMA~2 to create promising results, we concentrated on detecting illegality and improving diversity during the fine-tuning progress. 
LLM may distort or omit content in the input that renders the output unusable, such as excluding certain labels or slightly altering the transcript. Irrelevant hallucination and contextual continuation between the multiple inputs also needs to be prevented. Therefore, we conducted thorough verification to abandon the low-quality generation.
To better cultivate diversity and reduce the rough connection of labels instead of generating coherent sentences, three types of data enhancement methods were conducted, including order rearrangement of input attributes, expression synonym substitution, and multiple rounds of description translation.
GPT-4 Turbo~\cite{openai_gpt-4_2024} is used to undertake a primary proportion of data and the fine-tuned LLaMA~2 undertakes the main role of rewriting.

Remarkably, we aimed to generate two versions of speech prompts to make the data more widely applicable. 
The speech description (denoted as the \textbf{\textit{Description}} version) contains all available attributes regardless of the transcript. Besides, we involved speech transcript as an extra attribute to form a so-called speech instruction (denoted as the \textbf{\textit{Instruction}} version). 
The primary motivation for involving transcripts in the text prompt can be described as the convenience for unified control tendency in the future development of speech interpretation.
\begin{itemize}[noitemsep, topsep=0pt, partopsep=0pt, leftmargin=*]
    \item \textbf{Unified control of style and content.} Both style and transcript are significant in achieving unified control formation of speech audio in text speech interaction. 
    \item \textbf{Unified control of global and detailed style information.} Previous models process transcript and description through two channels respectively, resulting in difficulty to model the fine-grained instruction among different parts of transcript.
With the transcript contained in the description, it would be easier to model the global style, additionally, fine-grained instruction.
    \item \textbf{Unified control of speech and audio.} The unified construction form can be easily expanded from speech instruction to audio instruction datasets in the future, describing the overall scenario of sound events, speech content, and scene atmosphere.
\end{itemize}

\label{sect_express_speech_foundation_model}

\section{SpeechCraft Dataset}
\label{sec:pagestyle}
Deploying the annotation system on public speech datasets, here we introduce \textsc{SpeechCraft}, a bilingual dataset for fine-grained and expressive descriptions of speech. 
It is the largest open-sourced text description dataset in terms of both data scale and number of properties to characterize style.
\textsc{SpeechCraft} encompasses a total of 2,108,710 speech descriptions for approximately 1,000 hours of audio data per language.
Due to the absence of public data with emphasis labels, we propose a method to generate speech with stress pronunciation on word emphasis with paired description to cater to word-level fine-grained style control.
Furthermore, we conduct experiments to test the effectiveness of components within the data construction process in this section.
\subsection{Data Sources}

To facilitate further research in speech-language learning, we have implemented an annotation system for large-scale speech datasets, including precise attribute labeling and the crafting of expressive portrayals.
Considering a comprehensive set of factors such as dataset audio quality, the number and distribution of speakers, and the richness of emotional tones, we have selected the Chinese AISHELL-3 and Zhvoice, along with the English GigasSpeech-m and LibriTTS-R, as our foundational datasets.
Word clouds illustrating these datasets are presented in Fig. \ref{fig:teaser}. Detailed implementations of these four datasets are provided in the supplementary materials.





Notably, given the high-quality of corpus such as AISHELL-3, which explicitly marks various speaker attributes (gender, age group, native accents), some functionalities of the system were omitted in the actual construction of \textsc{SpeechCraft}, adapting to local conditions.

\subsection{Fine-grained Emphasis Speech Dataset}

\noindent\textbf{TTS Backbone with Disentangled Feature Control.}
Word emphasis is typically achieved through a combination of pitch variation, volume increase, elongation of sounds, and strategic pauses. 
These vocal cues work together to draw attention to the emphasized words. 
To better simulate the principles behind the generation of emphasis, we employed FastSpeech 2~\cite{ren_fastspeech_2020} as the backbone due to its ability to generate high-quality speech with disentangled control over phoneme-level acoustic properties, including pitch, energy, and duration.
By adjusting the predictions of these acoustic features, we can achieve increased volume, higher pitch, and prolonged sounds on the designed words. 
We tested a mix of energy, pitch, and duration scaling factors to pinpoint the optimal combination that aligns with human perception.\\

\noindent\textbf{Emphasis Speech Generation.}
Regarding the decision on emphasized words, we primarily assume that the keywords of a sentence are the most reasonable candidates for emphasis. Thus, we conducted keyword extraction from the transcripts of speech datasets.
The FastSpeech 2 model was pre-trained on AISHELL-3 and LibriTTS-R for the purpose of regenerating both datasets with keyword emphasis. Ultimately, we obtained 63,000 emphasized audio clips from AISHELL-3 and 75,000 from LibriTTS.
A detailed implementation of emphasis speech generation is illustrated in the supplementary materials.


\begin{figure}[t]
    \includegraphics[width=\columnwidth]{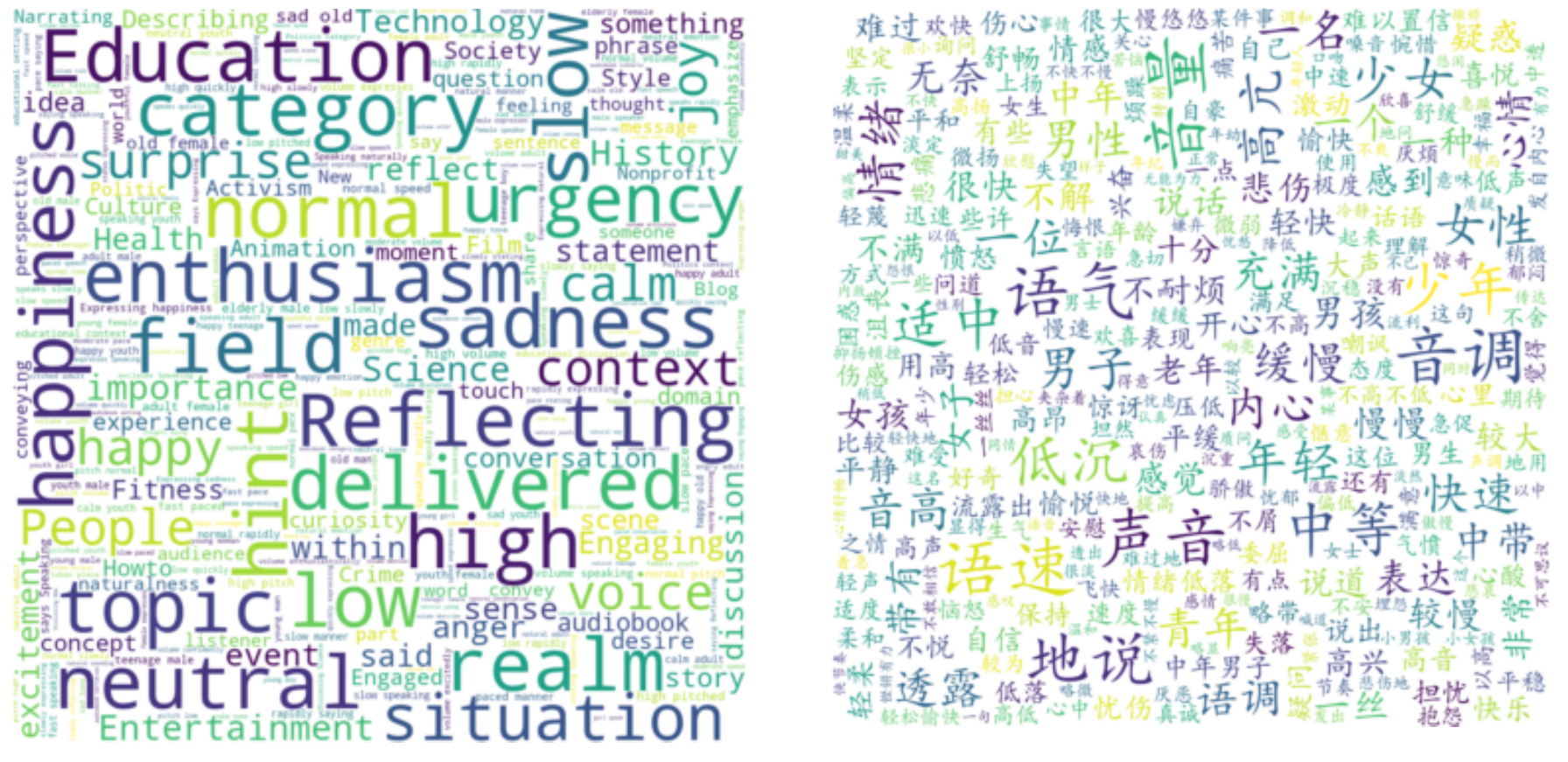}
    \vspace{-2mm}
    \begin{subfigure}[b]{0.45\columnwidth}
        \centering
        \caption{English}
    \end{subfigure}
    \begin{subfigure}[b]{0.45\columnwidth}
        \centering
        \caption{Chinese}
    \end{subfigure}
    \vspace{-1mm}
    \caption{Word clouds of top 300 words in English and Chinese parts of \textsc{SpeechCraft}. }
    \label{fig:teaser}
\end{figure}

\subsection{Validation of the Annotation System}
\begin{figure}[t]
  \includegraphics[width=0.95\columnwidth]{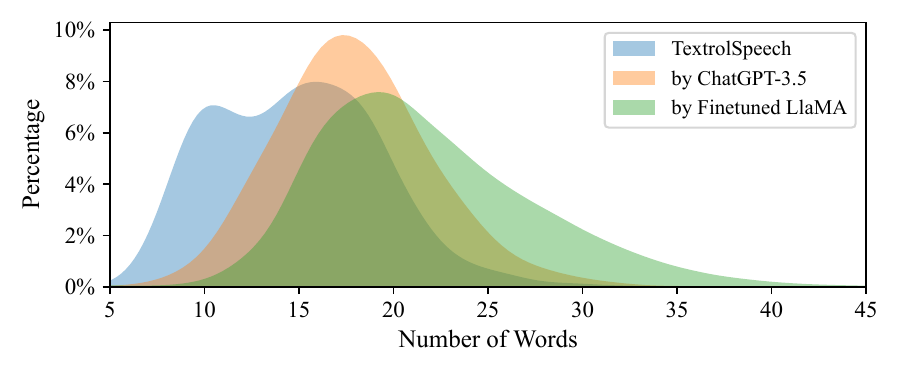}
  \vspace{-5pt}
  \caption{Sentence length distributions of the speech descriptions generated by different models.}
  \label{fig:Emotion}
\end{figure}

The overall annotation system works by integrating detailed audio analysis with language model rewriting to create descriptions that can capture the nuances of speech expressiveness automatically. 
As a result, it is important to validate the accuracy of detailed attributes and the performance of the fine-tuned LLM in overall rewriting.

Initially, we conduct an evaluation of the attribute predictors in style recognition using human-labeled data.
The Wav2vec~2.0 ~\cite{baevski_wav2vec_2020} based age and gender classification model achieved precisions of 97.72\% and 87.7\% respectively on the AISHELL~3 dataset, which provides authentic speaker information.
The officially released fine-tuned version of Emotion2vec~\cite{ma_emotion2vec_2023} model for speech emotion recognition task outperformed other SERs with an accuracy of 84\% on the internal English emotion dataset.
To evaluate the performance of SECap~\cite{xu_secap_2024}, which describes speech emotion with short sentences, we employed ChatGPT to summarize the main emotional tendency of the captioned sentence within the range of single-word emotions. The results indicated that the accuracy of the summary of SECap captions is 70.45\% on a twelve-class internal Chinese emotion dataset.

Subsequently, we implement a series of assessments on the annotations of the fine-tuned LLaMA~2 with multiple dimensions to evaluate the effect of rewriting. We take the descriptions produced by GPT-3.5 Turbo~\cite{brown_language_2020} and TextrolSpeech as baselines.
The \textbf{\textit{accuracy and completeness in preserving given labels}} reflect the fidelity of the language model in preserving authentic information through the rewriting process. 
As shown in Tab.~\ref{tab:assessment}, the LLaMA~2 with instructional fine-tuning on data generated by GPT~4.0 had competitive results with GPT-3.5 Turbo~\cite{brown_language_2020}. 
The overall error rates of the fine-tuned LLaMA was the lowest among three types of descriptions, indicating less omission or distortion throughout the transformation process. It thereby guaranteed precise semantics embedded within a broader range of intricate attributes.
The \textbf{\textit{distribution of sentence length}} in descriptions indirectly indicates that our longer sentences possess the potential to encapsulate a greater depth of detail, as shown in Fig. \ref{fig:Emotion}.
Another interesting statistical discovery concerns the \textbf{\textit{position of spoken transcript among the speech instruction annotation}}. The transcript exhibits probabilities of 5.55\%, 33.05\%, and 61.40\% for occurring at the beginning, middle, and end of a sentence, respectively, showcasing syntactic diversity.

\newlength{\nearlyquartercolumnwidth}
\setlength{\nearlyquartercolumnwidth}{0.22\columnwidth}
\begin{table}[h]
\centering
\caption{Assessment on rewriting of the fine-tuned LLaMA~2.}
\resizebox{\columnwidth}{!}{
\begin{tabular}{@{}cccccC{1.2cm}C{1.2cm}@{}}
\toprule
\multirow{2}{*}{\textbf{Prompt}} & \multicolumn{1}{c}{\textbf{TextrolSpeech}} & & \multicolumn{1}{c}{\textbf{GPT-3.5 Turbo}} & & \multicolumn{2}{c}{\textbf{Fintuned LLaMA~2}} \\ 
\cline{2-2} \cline{4-4} \cline{6-7}
& \multicolumn{1}{c}{\textbf{EN}} & & \multicolumn{1}{c}{\textbf{ZH}} & & \multicolumn{1}{c}{\textbf{EN}} & \multicolumn{1}{c}{\textbf{ZH}} \\ 
\midrule
\textbf{Omission} & 14.80\% & & 4.04\% & &  1.95\% & 3.88\%\\ 
\textbf{Distortion} & \textcolor{white}{0}2.10\% & & 6.10\% & & 7.50\% & 6.14\%\\ 
\cdashline{2-2} \cdashline{4-4} \cdashline{6-7}
\textbf{MOS} & 3.58 & & \multicolumn{1}{c}{3.84} & & \multicolumn{2}{c}{4.02}\\
\bottomrule
\end{tabular}
}
\label{tab:assessment}
\end{table}


\begin{table}[!htbp]
\setcounter{oldfootnote}{\value{footnote}}
\setcounter{footnote}{0}
\renewcommand{\thefootnote}{\fnsymbol{footnote}}
\centering
\caption{Data source distribution in \textsc{SpeechCraft}}
\vspace{-5pt}
\label{tab:datasets scale Comparison}
\resizebox{0.95\columnwidth}{!}{
\begin{tabular}{@{}C{2.1cm}C{\nearlyquartercolumnwidth}R{\nearlyquartercolumnwidth}R{\nearlyquartercolumnwidth}@{}}
\toprule
\textbf{Dataset} & \textbf{Language} &\textbf{\#Duration} & \textbf{\#Clips}\\
\midrule
\textsc{SpeechCraft} & EN + ZH & 2,381.54h & 2,249,579\\
\hdashline 
LibriTTS-R\footnotemark[2] & EN & 697.66h & 427,919\\
GigaSpeech-m & EN & 739.91h & 670,070\\
AISHELL-3\footnotemark[2] & ZH & 114.29h & 126,520\\
Zhvoice & ZH & 799.68h & 1,025,070\\
\bottomrule
\end{tabular}
}
\begin{flushleft}
    \footnotesize
    \quad\footnotemark[2]~Statistics contain the original version and regeneratation with emphasis.
\end{flushleft}
\setcounter{footnote}{\value{oldfootnote}}
\renewcommand{\thefootnote}{\arabic{footnote}}
\end{table}

\begin{figure}[h]
  \includegraphics[width=0.975\columnwidth]{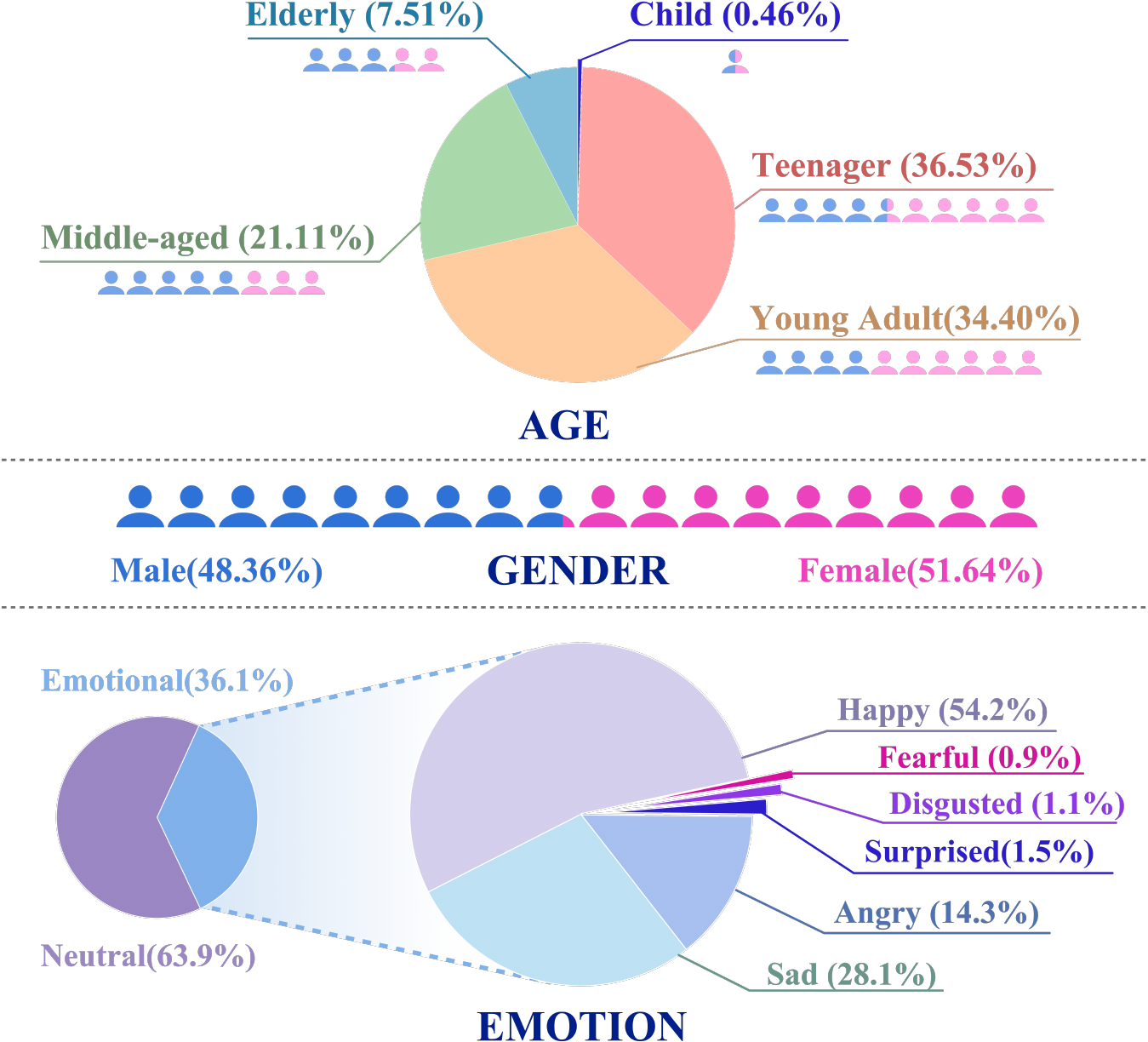}
  \vspace{-5pt}
  \caption{Distributions of age, gender and emotion.}
  \label{fig:AgeGenderEmotion}
\end{figure}

\begin{table*}[t]
\centering
\caption{Experimental results of expressive speech synthesis with TextrolSpeech and \textsc{SpeechCraft}.}
\vspace{-5pt}
\renewcommand{\thefootnote}{\fnsymbol{footnote}}
\resizebox{\textwidth}{!}{
\begin{tabular}{@{}C{1.46cm}C{2.0cm}C{1.2cm}C{1.2cm}cC{1.1cm}C{0.9cm}C{0.9cm}C{0.9cm}C{1.1cm}C{1.5cm}C{1.5cm}C{1cm}@{}}
\toprule
\multirow{2}{*}{\textbf{Method}} & \multirow{2}{*}{\textbf{Dataset}} & \multicolumn{2}{c}{\textbf{TTS Quality}} & & \multicolumn{7}{c}{\textbf{Acc on Style Factors}} & \multirow{2}{*}{\textbf{MOS}} \\
\cline{3-4} \cline{6-12}
 & & \textbf{MCD}$\downarrow$ & \textbf{SECS}$\uparrow$ & & \textbf{Gender} & \textbf{Age}\footnotemark[1] & \textbf{Pitch} & \textbf{Speed} & \textbf{Energy} & \textbf{Emotion} & \textbf{Mean\_Acc}& \\
\midrule
- & GroundTruth & - & - & & 100 & 100 & 82.03  & 90.52 & 80.39 & 100 & 92.16 & - \\
 \cdashline{1-13}\multirow{2}{*}{\textbf{Salle}} & TextrolSpeech & 15.26 & 56.10 & & 78.16 & 47.78 & 49.05 & 56.91 & 51.27 & 53.48 & 56.11 & 2.12\\
 & \textsc{Ours} (Des) & \textbf{12.87} & 60.90 & & \textbf{94.30} & 78.48 & 60.76 & 62.97 & 61.71 & 59.30 & 69.59 & 4.23 \\
 \cdashline{1-13}
\multirow{3}{*}{\textbf{ParlerTTS}} & TextrolSpeech & 21.94 & 54.80 & & 87.14 & 61.74 & 55.81 & 58.80 & 63.12 & 66.11 & 75.62 & 3.52\\
 & \textsc{Ours} (Des) & 13.22 & \textbf{61.70} & & 92.60 & \textbf{87.46} & 65.92  & 71.70 & 83.60 & \textbf{81.99} & 80.54 & \textbf{4.56} \\
& \textsc{Ours} (Ins) & 16.98 & 60.50 & & 94.02 & 85.21 & \textbf{68.77}  & \textbf{72.35} & \textbf{83.92} & 79.10 & \textbf{80.56} & 4.43 \\
\bottomrule
\end{tabular}
}
\label{sec:TTS Result}
\begin{flushleft}
\footnotesize
\quad\footnotemark[1] TextrolSpeech dataset doesn't contain age label. 
\end{flushleft}
\end{table*}

Finally, we conducted a mutual experiment to evaluate both the quality of the regenerated emphasis data and the effectiveness of the proposed emphasis detection model.
Initially, we trained models using the proposed regeneration of AISHELL-3 and LibriTTS-R datasets respectively. The AISHELL-3-stressed achieved an accuracy of 88.55\% on test set, while the LibriTTS-R-stressed achieves an accuracy of  85.60\%.
Additionally, the emphasis experiment on an internal dataset, with human annotation on word emphasis, is provided in supplementary materials, further demonstrating the effectiveness of our approach in modeling real-life stress patterns.


\subsection{Data Analysis}

The \textsc{SpeechCraft} dataset is distributed evenly across both Chinese and English datasets, encompassing over 2,000,000 audio clips annotated with speech \textbf{\textit{Descriptions}} and speech \textbf{\textit{Instructions}}. 
The detailed information of the modified data in \textsc{SpeechCraft} that sourced from public speech datasets are listed in Tab.~\ref{tab:datasets scale Comparison}.

Distribution of gender and age are shown in Fig.~\ref{fig:AgeGenderEmotion}. 
The gender distribution is nearly balanced with a slight variation among different age groups. 
Most ages are distributed as \textsc{Teenager}, \textsc{Young Adult}, and \textsc{Middle-aged}, indicating a focus on individuals likely within the most active stages of life.
English part of \textsc{SpeechCraft} contains 36.1\% emotional data mostly regarded as \textsc{Happy}, \textsc{Sad}, and \textsc{Angry}. 
The quantities of clips expressing \textsc{Surprised}, \textsc{Disgusted}, and \textsc{Fearful} are 5,626, 4,038, and 3,223, respectively, exceeding the scale of existing datasets. However, their proportions remain low in the context of the vast total amount of data.
The unbalanced distribution of emotions resulting from in-the-wild data also indicates the frequency of real-life emotion tendency outside datasets. 




\section{Boost Performance via SpeechCraft}
\label{sect_experi}

To assess the impact of the proposed \textsc{SpeechCraft} dataset, we conduct comprehensive experiments across various speech-language multi-modal learning tasks, including expressive speech synthesis, fine-grained emphasis control in TTS systems, and automated speech style captioning.

\subsection{Expressive Speech Synthesis}

The expressive speech synthesis task aims to generate high-quality speech audio under the intended speaking style in a seamless manner.
Typically, the task employs a natural language description prompt as input to modulate the expressiveness of style.
To validate the effectiveness of \textsc{SpeechCraft} in replicating the intended expressiveness, we reproduce two text prompt TTS models, Salle and ParlerTTS~\cite{lyth2024natural}, and conduct training on both the TextrolSpeech dataset and the proposed \textsc{SpeechCraft} dataset for comparative analysis.
Both Salle and ParlerTTS facilitate speech synthesis task by conditional codec language modeling with residual vector quantization ~\cite{défossez2022highfidelityneuralaudio} (RVQ) as audio representations, but mainly differs in their methods of autoregressive token generation.

As the TextrolSpeech dataset does not contain transcripts within its descriptions, we perform rigorous comparison experiments on the \textbf{\textit{Description}} version of \textsc{SpeechCraft}. 
We strictly follow the official guidelines, training each model 600,000 steps on Salle and 50,000 steps on ParlerTTS.
The evaluation was conducted across three key dimensions: the recall accuracy of the style factors, audio quality, and user study.
The testset is consist of 316 randomly sampled audio clips from GigaSpeech-s TTS corpus, embodying a diverse array of all attribute dimensions.
We used the speech annotation pipeline in Sec.~\ref{sec:Speech Style Recognition} to characterize the groundtruth and synthesized speech. Attributes recall accuracy was calculated by comparing the two versions of style labels.
Traditional objective evaluation metrics, namely Speaker Encoder Cosine Similarity
(SECS~\cite{casanova2021sc}) and Mel-Cepstral Distortion (MCD), were further utilized to assess the similarity of speaker identity and MFCC features between synthesized and original speech. 

As illustrated in Tab.~\ref{sec:TTS Result}, the substantial potential utility of large scale data is evidenced by the English subset of \textsc{SpeechCraft} \textbf{\textit{Description}} outperforming TextrolSpeech from all aspects and the mean accuracy with ParlerTTS reached 80\%.
Notably, the labels of pitch, energy and speed are classified based on a relative percentage borderline within the testset domain.


\subsection{Fine-grained Speech Emphasis Control}

\begin{table*}[t]
\centering
\caption{Experimental results of automated speech captioning on the test set of SECap. Each type of highlight/underline represents a unique speech attribute. In addition to the existing emotion attribute, speech style captioning with \textsc{SpeechCraft} captured properties of age, gender, pitch, energy, speed for the first time.}
\vspace{-5pt}
\resizebox{\textwidth}{!}{
\begin{tabular}
{@{}C{6.5cm}C{11.5cm}@{}}
\toprule
\textbf{SECap} & \textbf{Captioning trained on \textsc{SpeechCraft}}\\
\midrule
Felt \colorbox{emotion}{happiness and joy}.  & A \colorbox{age}{young} \colorbox{gender}{woman}, \colorbox{energy}{voice high}, \colorbox{speed}{pace swift}, revealed \colorbox{emotion}{joy and delight} in her emotion.\\
\hdashline
Appears to be very skillful. & A \colorbox{age}{young} \colorbox{gender}{gentleman}, with an \colorbox{pitch}{elevated pitch} and \colorbox{speed}{rapid speed}, articulated in \colorbox{emotion}{anger}.\\
\hdashline
The voice was full of \colorbox{emotion}{curiosity}, and the tone carried a careful \colorbox{emotion}{anticipation}.  & A \colorbox{age}{young} \colorbox{gender}{female}, with a \colorbox{pitch}{high-pitched} voice and a \colorbox{speed}{moderate pace}, spoke with an air of \colorbox{emotion}{confusion and misunderstanding}.\\
\hdashline
\colorbox{emotion}{Suspicious and puzzled} about something.  & A \colorbox{age}{young} \colorbox{gender}{female}'s tone was \colorbox{pitch}{high-pitched} and the \colorbox{speed}{pace was moderate}, speaking with a sense of \colorbox{emotion}{doubt}.\\
\bottomrule
\end{tabular}
}
\label{sec:Captioning Result}
\end{table*}

\begin{table}[!htbp]
\centering
\caption{The recall accuracy of emphasis detection. Abbr.: GT (Ground-truth), Des (\textbf{\textit{Description}}), Ins (\textbf{\textit{Instruction}}).}
\vspace{-5pt}
\resizebox{0.95\columnwidth}{!}{
\begin{tabularx}{0.95\columnwidth}{@{}cccccc@{}}
\toprule
\textbf{Lan} & \textbf{Version} & \textbf{$\text{Acc}_w$} & \textbf{$\text{Acc}_s$ \textit{R}@1} & \textbf{$\text{Acc}_s$ \textit{R}@2} & \textbf{MOS}\\
\midrule
\multirow{3}{*}{\textbf{EN}} & GT & 91.17 & 62.15 & 62.58 & 3.16 \\
& Des & 76.55 & 59.14 & 69.89 & 2.70 \\
& Ins & \textbf{88.53} & \textbf{87.77} & \textbf{90.96} & \textbf{3.98} \\
\hdashline
\multirow{3}{*}{\textbf{ZH}} 
& GT & 86.97 & 63.60 & 64.25 & 3.79 \\
 & Des & 64.83 & 89.31 & 91.60 & 2.84 \\
 & Ins & \textbf{84.02} & \textbf{94.06} & \textbf{95.43} & \textbf{4.05}\\
\bottomrule
\end{tabularx}
}
\label{recall stress}
\end{table}


Utilizing synthesized emphasis data with paired fine-grained style instructions, we explore the potential applications of speech emphasis control by fine-tuning the expressive speech synthesis model. 
To assess the impact of incorporating transcript into the text description on fine-grained control capabilities like emphasis, we compare the emphasized speech effects between the \textbf{\textit{Description}} version and the \textbf{\textit{Instruction}} version of \textsc{SpeechCraft}.

\begin{figure}[t]
  \includegraphics[width=0.9\columnwidth]{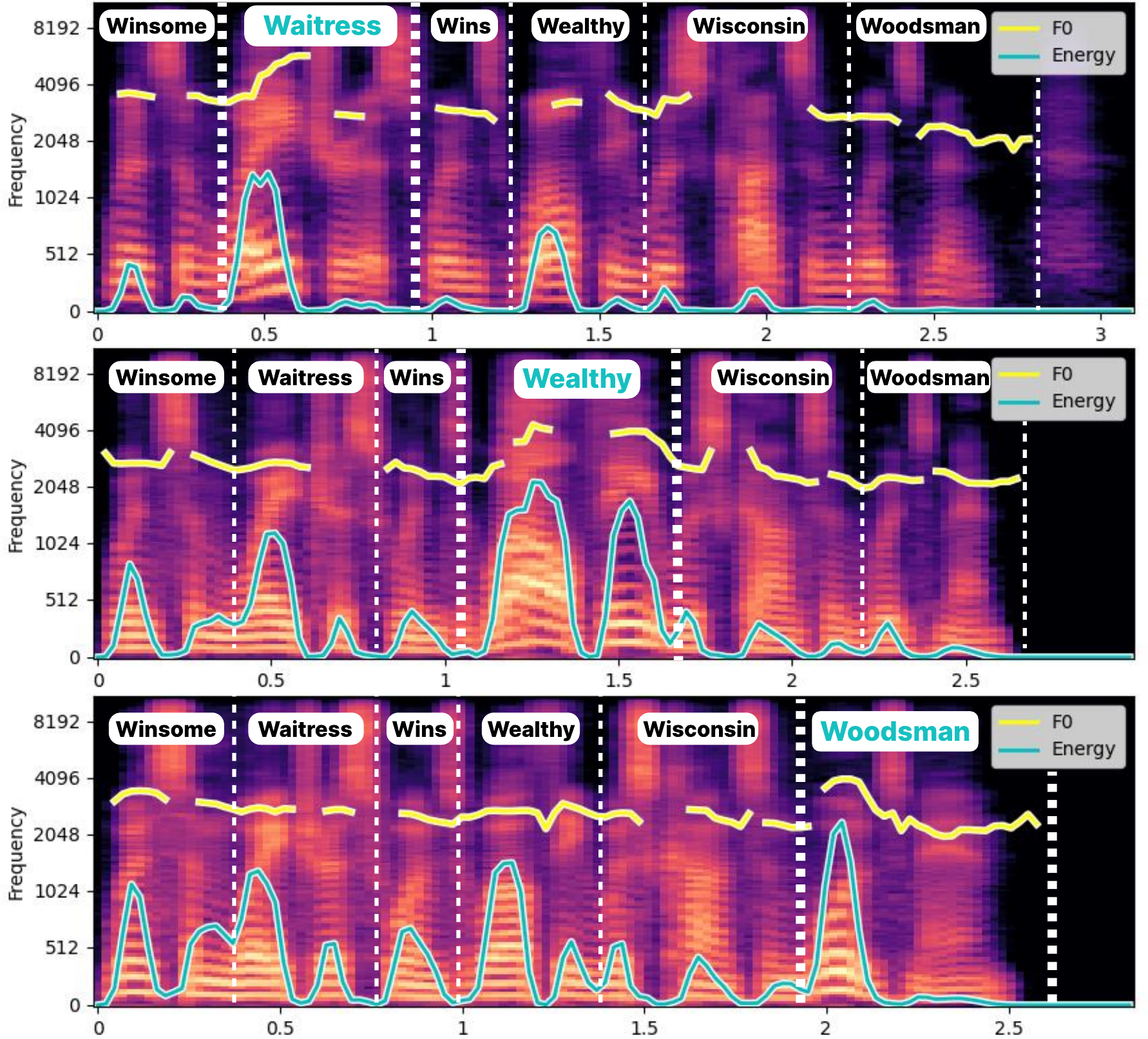}
  \vspace{-8pt}
  \caption{Mel-spectrogram examples of speech emphasis control. Different stressed words on `Waitress', `Wealthy' and `Woodsman' respectively with the same text instructions.}
  \label{fig:Mel}
\end{figure}

We evaluate emphasis accuracy in the synthesized speech in English and Chinese respectively. According to the design of the emphasis detection model in Sec.\ref{sec:Word Emphasis Detection}, emphasis is detected by the unit of words. 
Therefore, word-level accuracy ($\text{Acc}_w$) denotes the nominal accuracy across all word pieces segmented from sentences. Sentence-level accuracy ($\text{Acc}_s$) refers to the true accuracy of correctly predicting an emphasized word based on the complete sentence.
Importantly, we include instructions without emphasis demands in the test set to evaluate the system's reliability through precise emphasis control. Sentences without emphasis demands are considered correct only if no word is incorrectly identified as emphasized.
As illustrated in Tab.\ref{recall stress}, the proposed \textbf{\textit{Instruction}} version excels over the \textbf{\textit{Description}} version across all metrics, achieving 95.43\% accuracy in Chinese sentences and 90.96\% in English. Furthermore, ablation experiments on ParlerTTS in Tab.\ref{sec:TTS Result} indicates that involving transcript does not decrease the performance of other global style factors.
This demonstrates conclusively that our data, enriched with emphasis word instructions and specific features of stressed pronunciation, fuses the fine-grained speech emphasis control with outstanding effectiveness.

\noindent \textbf{Case Study.} We further conduct a case study using a series of same base instructions varied only in the words emphasized, as illustrated in Fig.~\ref{fig:Mel}. The line of speech energy (green) shows a clear peak at the targeted words and aligns with the highest fundamental frequency (yellow) within the sentence. The distinct emphasis on the words highlights the flexibility and precision of our dataset's fine-grained, style-controllable synthesis capabilities.

\vspace{-1mm}
\subsection{Automated Speech Style Captioning}
\label{sct:automated_speech_style_captioning}


Automated speech style captioning transcends emotion captioning by providing comprehensive descriptions that encompass not only the emotional tone but also stylistic nuances such as acoustic properties and speaker identity.
Owing to the nascent stage of speech style captioning development, we replicate the state-of-the-art automated emotion captioning model SECap (refer to Sec.~\ref{sect_expressive_tts}), and retrain it on the \textbf{\textit{Description}} version of \textsc{SpeechCraft}. Utilizing the semantic features extracted by the pre-trained audio encoder within SECap, the model achieves a remarkably coherent interpretation of the overall speech style to a significant degree.
For the first time, it was capable of capturing a descriptive sentence that included acoustic properties, speaker identity, and emotional tone. This accomplishment marks a substantial advancement in the nuanced articulation of speech characteristics. 
Some of the cases are documented in Tab.~\ref{sec:Captioning Result}.

We conduct a user study to evaluate the detail and accuracy of the descriptions in capturing the audio style. Although our model marginally outperforms the original SECap (3.79 vs 3.58), there is consensus that speech style captioning, while capturing a broader range of audio characteristics, does so in a less expressive manner compared to SECap. SECap deals with nuanced emotional variations more explicitly, highlighting the need for further research into the trade-offs involved in fine-grained speech style captioning regarding the scope and detail of dimensions.

\vspace{-1mm}
\section{Conclusion}
\label{sect_concl}


In this work, we proposed an automatic speech annotation system for expressiveness interpretation, which adopted refined speech style recognition with LLMs rewriting to form detailed and customized speech descriptions.
Furthermore, we created \textsc{SpeechCraft}, the largest open-source expressive bilingual speech dataset with natural language descriptions, which has great potential in large speech-language model training.
Experiments showed that our dataset strongly enhances the performance of both text-to-speech and speech-to-text models. 
It boosts fine-grained style control such as word emphasis in expressive speech synthesis, and facilitates automated speech-style captioning in the broader field of speech comprehension and interaction technologies.

\begin{acks}
This work is supported by the National Key R\&D Program of China under Grant No.2024QY1400, National Natural Science Foundation of China under Grant No.62076144 and Shenzhen Science and Technology Program under Grant No.WDZC20220816140515001.
We would like to thank Zhipu AI for their valuable support.
\end{acks}



\bibliographystyle{ACM-Reference-Format}
\balance
\bibliography{paper/reference.bib}


\begin{thebibliography}{52}


\ifx \showCODEN    \undefined \def \showCODEN     #1{\unskip}     \fi
\ifx \showDOI      \undefined \def \showDOI       #1{#1}\fi
\ifx \showISBNx    \undefined \def \showISBNx     #1{\unskip}     \fi
\ifx \showISBNxiii \undefined \def \showISBNxiii  #1{\unskip}     \fi
\ifx \showISSN     \undefined \def \showISSN      #1{\unskip}     \fi
\ifx \showLCCN     \undefined \def \showLCCN      #1{\unskip}     \fi
\ifx \shownote     \undefined \def \shownote      #1{#1}          \fi
\ifx \showarticletitle \undefined \def \showarticletitle #1{#1}   \fi
\ifx \showURL      \undefined \def \showURL       {\relax}        \fi
\providecommand\bibfield[2]{#2}
\providecommand\bibinfo[2]{#2}
\providecommand\natexlab[1]{#1}
\providecommand\showeprint[2][]{arXiv:#2}

\bibitem[Ardila et~al\mbox{.}(2019)]%
        {ardila2019common}
\bibfield{author}{\bibinfo{person}{Rosana Ardila}, \bibinfo{person}{Megan Branson}, \bibinfo{person}{Kelly Davis}, \bibinfo{person}{Michael Henretty}, \bibinfo{person}{Michael Kohler}, \bibinfo{person}{Josh Meyer}, \bibinfo{person}{Reuben Morais}, \bibinfo{person}{Lindsay Saunders}, \bibinfo{person}{Francis~M Tyers}, {and} \bibinfo{person}{Gregor Weber}.} \bibinfo{year}{2019}\natexlab{}.
\newblock \showarticletitle{Common voice: A massively-multilingual speech corpus}.
\newblock \bibinfo{journal}{\emph{arXiv preprint arXiv:1912.06670}} (\bibinfo{year}{2019}).
\newblock


\bibitem[Baevski et~al\mbox{.}(2020)]%
        {baevski_wav2vec_2020}
\bibfield{author}{\bibinfo{person}{Alexei Baevski}, \bibinfo{person}{Yuhao Zhou}, \bibinfo{person}{Abdelrahman Mohamed}, {and} \bibinfo{person}{Michael Auli}.} \bibinfo{year}{2020}\natexlab{}.
\newblock \showarticletitle{wav2vec 2.0: {A} {Framework} for {Self}-{Supervised} {Learning} of {Speech} {Representations}}. In \bibinfo{booktitle}{\emph{Advances in {Neural} {Information} {Processing} {Systems}}}, Vol.~\bibinfo{volume}{33}. \bibinfo{publisher}{Curran Associates, Inc.}, \bibinfo{pages}{12449--12460}.
\newblock
\urldef\tempurl%
\url{https://proceedings.neurips.cc/paper_files/paper/2020/file/92d1e1eb1cd6f9fba3227870bb6d7f07-Paper.pdf}
\showURL{%
\tempurl}


\bibitem[Bai et~al\mbox{.}(2023)]%
        {bai_qwen_2023}
\bibfield{author}{\bibinfo{person}{Jinze Bai}, \bibinfo{person}{Shuai Bai}, \bibinfo{person}{Yunfei Chu}, \bibinfo{person}{Zeyu Cui}, \bibinfo{person}{Kai Dang}, \bibinfo{person}{Xiaodong Deng}, \bibinfo{person}{Yang Fan}, \bibinfo{person}{Wenbin Ge}, \bibinfo{person}{Yu Han}, \bibinfo{person}{Fei Huang}, {and} \bibinfo{person}{{others}}.} \bibinfo{year}{2023}\natexlab{}.
\newblock \showarticletitle{Qwen technical report}.
\newblock \bibinfo{journal}{\emph{arXiv preprint arXiv:2309.16609}} (\bibinfo{year}{2023}).
\newblock


\bibitem[Brown et~al\mbox{.}(2020)]%
        {brown_language_2020}
\bibfield{author}{\bibinfo{person}{Tom Brown}, \bibinfo{person}{Benjamin Mann}, \bibinfo{person}{Nick Ryder}, \bibinfo{person}{Melanie Subbiah}, \bibinfo{person}{Jared~D Kaplan}, \bibinfo{person}{Prafulla Dhariwal}, \bibinfo{person}{Arvind Neelakantan}, \bibinfo{person}{Pranav Shyam}, \bibinfo{person}{Girish Sastry}, \bibinfo{person}{Amanda Askell}, \bibinfo{person}{Sandhini Agarwal}, \bibinfo{person}{Ariel Herbert-Voss}, \bibinfo{person}{Gretchen Krueger}, \bibinfo{person}{Tom Henighan}, \bibinfo{person}{Rewon Child}, \bibinfo{person}{Aditya Ramesh}, \bibinfo{person}{Daniel Ziegler}, \bibinfo{person}{Jeffrey Wu}, \bibinfo{person}{Clemens Winter}, \bibinfo{person}{Chris Hesse}, \bibinfo{person}{Mark Chen}, \bibinfo{person}{Eric Sigler}, \bibinfo{person}{Mateusz Litwin}, \bibinfo{person}{Scott Gray}, \bibinfo{person}{Benjamin Chess}, \bibinfo{person}{Jack Clark}, \bibinfo{person}{Christopher Berner}, \bibinfo{person}{Sam McCandlish}, \bibinfo{person}{Alec Radford}, \bibinfo{person}{Ilya Sutskever}, {and}
  \bibinfo{person}{Dario Amodei}.} \bibinfo{year}{2020}\natexlab{}.
\newblock \showarticletitle{Language {Models} are {Few}-{Shot} {Learners}}. In \bibinfo{booktitle}{\emph{Advances in {Neural} {Information} {Processing} {Systems}}}, Vol.~\bibinfo{volume}{33}. \bibinfo{publisher}{Curran Associates, Inc.}, \bibinfo{pages}{1877--1901}.
\newblock
\urldef\tempurl%
\url{https://proceedings.neurips.cc/paper_files/paper/2020/file/1457c0d6bfcb4967418bfb8ac142f64a-Paper.pdf}
\showURL{%
\tempurl}


\bibitem[Busso et~al\mbox{.}(2008)]%
        {busso_iemocap_2008}
\bibfield{author}{\bibinfo{person}{Carlos Busso}, \bibinfo{person}{Murtaza Bulut}, \bibinfo{person}{Chi-Chun Lee}, \bibinfo{person}{Abe Kazemzadeh}, \bibinfo{person}{Emily Mower}, \bibinfo{person}{Samuel Kim}, \bibinfo{person}{Jeannette~N. Chang}, \bibinfo{person}{Sungbok Lee}, {and} \bibinfo{person}{Shrikanth~S. Narayanan}.} \bibinfo{year}{2008}\natexlab{}.
\newblock \showarticletitle{{IEMOCAP}: interactive emotional dyadic motion capture database}.
\newblock \bibinfo{journal}{\emph{Language Resources and Evaluation}} \bibinfo{volume}{42}, \bibinfo{number}{4} (\bibinfo{date}{Dec.} \bibinfo{year}{2008}), \bibinfo{pages}{335--359}.
\newblock
\showISSN{1574-0218}
\urldef\tempurl%
\url{https://doi.org/10.1007/s10579-008-9076-6}
\showDOI{\tempurl}


\bibitem[Casanova et~al\mbox{.}(2021)]%
        {casanova2021sc}
\bibfield{author}{\bibinfo{person}{Edresson Casanova}, \bibinfo{person}{Christopher Shulby}, \bibinfo{person}{Eren G{\"o}lge}, \bibinfo{person}{Nicolas~Michael M{\"u}ller}, \bibinfo{person}{Frederico~Santos De~Oliveira}, \bibinfo{person}{Arnaldo~Candido Junior}, \bibinfo{person}{Anderson da~Silva Soares}, \bibinfo{person}{Sandra~Maria Aluisio}, {and} \bibinfo{person}{Moacir~Antonelli Ponti}.} \bibinfo{year}{2021}\natexlab{}.
\newblock \showarticletitle{SC-GlowTTS: An efficient zero-shot multi-speaker text-to-speech model}.
\newblock \bibinfo{journal}{\emph{arXiv preprint arXiv:2104.05557}} (\bibinfo{year}{2021}).
\newblock


\bibitem[Chan and Chin(2020)]%
        {chan_comprehensive_2020}
\bibfield{author}{\bibinfo{person}{T.~K. Chan} {and} \bibinfo{person}{Cheng~Siong Chin}.} \bibinfo{year}{2020}\natexlab{}.
\newblock \showarticletitle{A {Comprehensive} {Review} of {Polyphonic} {Sound} {Event} {Detection}}.
\newblock \bibinfo{journal}{\emph{IEEE Access}}  \bibinfo{volume}{8} (\bibinfo{year}{2020}), \bibinfo{pages}{103339--103373}.
\newblock
\urldef\tempurl%
\url{https://doi.org/10.1109/ACCESS.2020.2999388}
\showDOI{\tempurl}


\bibitem[Chen et~al\mbox{.}(2021)]%
        {chen_gigaspeech_2021}
\bibfield{author}{\bibinfo{person}{Guoguo Chen}, \bibinfo{person}{Shuzhou Chai}, \bibinfo{person}{Guanbo Wang}, \bibinfo{person}{Jiayu Du}, \bibinfo{person}{Wei-Qiang Zhang}, \bibinfo{person}{Chao Weng}, \bibinfo{person}{Dan Su}, \bibinfo{person}{Daniel Povey}, \bibinfo{person}{Jan Trmal}, \bibinfo{person}{Junbo Zhang}, {and} \bibinfo{person}{{others}}.} \bibinfo{year}{2021}\natexlab{}.
\newblock \showarticletitle{Gigaspeech: {An} evolving, multi-domain asr corpus with 10,000 hours of transcribed audio}.
\newblock \bibinfo{journal}{\emph{arXiv preprint arXiv:2106.06909}} (\bibinfo{year}{2021}).
\newblock


\bibitem[Drossos et~al\mbox{.}(2020)]%
        {drossos_clotho_2020}
\bibfield{author}{\bibinfo{person}{Konstantinos Drossos}, \bibinfo{person}{Samuel Lipping}, {and} \bibinfo{person}{Tuomas Virtanen}.} \bibinfo{year}{2020}\natexlab{}.
\newblock \showarticletitle{Clotho: an {Audio} {Captioning} {Dataset}}. In \bibinfo{booktitle}{\emph{{ICASSP} 2020 - 2020 {IEEE} {International} {Conference} on {Acoustics}, {Speech} and {Signal} {Processing} ({ICASSP})}}. \bibinfo{pages}{736--740}.
\newblock
\urldef\tempurl%
\url{https://doi.org/10.1109/ICASSP40776.2020.9052990}
\showDOI{\tempurl}


\bibitem[Dupuis and Pichora-Fuller(2010)]%
        {dupuis_toronto_2010}
\bibfield{author}{\bibinfo{person}{Kate Dupuis} {and} \bibinfo{person}{M~Kathleen Pichora-Fuller}.} \bibinfo{year}{2010}\natexlab{}.
\newblock \showarticletitle{Toronto emotional speech set (tess)-younger talker\_happy}.
\newblock  (\bibinfo{year}{2010}).
\newblock
\newblock
\shownote{Publisher: Toronto: University of Toronto, Psychology Department, 2010.}.


\bibitem[Défossez et~al\mbox{.}(2022)]%
        {défossez2022highfidelityneuralaudio}
\bibfield{author}{\bibinfo{person}{Alexandre Défossez}, \bibinfo{person}{Jade Copet}, \bibinfo{person}{Gabriel Synnaeve}, {and} \bibinfo{person}{Yossi Adi}.} \bibinfo{year}{2022}\natexlab{}.
\newblock \showarticletitle{High Fidelity Neural Audio Compression}.
\newblock \bibinfo{journal}{\emph{arXiv preprint arXiv:2210.13438}} (\bibinfo{year}{2022}).
\newblock


\bibitem[Guan et~al\mbox{.}(2024)]%
        {guan_mm-tts_2024}
\bibfield{author}{\bibinfo{person}{Wenhao Guan}, \bibinfo{person}{Yishuang Li}, \bibinfo{person}{Tao Li}, \bibinfo{person}{Hukai Huang}, \bibinfo{person}{Feng Wang}, \bibinfo{person}{Jiayan Lin}, \bibinfo{person}{Lingyan Huang}, \bibinfo{person}{Lin Li}, {and} \bibinfo{person}{Qingyang Hong}.} \bibinfo{year}{2024}\natexlab{}.
\newblock \showarticletitle{{MM}-{TTS}: {Multi}-{Modal} {Prompt} {Based} {Style} {Transfer} for {Expressive} {Text}-to-{Speech} {Synthesis}}.
\newblock \bibinfo{journal}{\emph{Proceedings of the AAAI Conference on Artificial Intelligence}} \bibinfo{volume}{38}, \bibinfo{number}{16} (\bibinfo{date}{March} \bibinfo{year}{2024}), \bibinfo{pages}{18117--18125}.
\newblock
\urldef\tempurl%
\url{https://doi.org/10.1609/aaai.v38i16.29769}
\showDOI{\tempurl}


\bibitem[Guo et~al\mbox{.}(2023)]%
        {guo_prompttts_2023}
\bibfield{author}{\bibinfo{person}{Zhifang Guo}, \bibinfo{person}{Yichong Leng}, \bibinfo{person}{Yihan Wu}, \bibinfo{person}{Sheng Zhao}, {and} \bibinfo{person}{Xu Tan}.} \bibinfo{year}{2023}\natexlab{}.
\newblock \showarticletitle{Prompttts: {Controllable} {Text}-{To}-{Speech} {With} {Text} {Descriptions}}. In \bibinfo{booktitle}{\emph{{ICASSP} 2023 - 2023 {IEEE} {International} {Conference} on {Acoustics}, {Speech} and {Signal} {Processing} ({ICASSP})}}. \bibinfo{pages}{1--5}.
\newblock
\urldef\tempurl%
\url{https://doi.org/10.1109/ICASSP49357.2023.10096285}
\showDOI{\tempurl}


\bibitem[Guzhov et~al\mbox{.}(2022)]%
        {guzhov_audioclip_2022}
\bibfield{author}{\bibinfo{person}{Andrey Guzhov}, \bibinfo{person}{Federico Raue}, \bibinfo{person}{Jörn Hees}, {and} \bibinfo{person}{Andreas Dengel}.} \bibinfo{year}{2022}\natexlab{}.
\newblock \showarticletitle{Audioclip: {Extending} {Clip} to {Image}, {Text} and {Audio}}. In \bibinfo{booktitle}{\emph{{ICASSP} 2022 - 2022 {IEEE} {International} {Conference} on {Acoustics}, {Speech} and {Signal} {Processing} ({ICASSP})}}. \bibinfo{pages}{976--980}.
\newblock
\urldef\tempurl%
\url{https://doi.org/10.1109/ICASSP43922.2022.9747631}
\showDOI{\tempurl}


\bibitem[Ji et~al\mbox{.}(2024)]%
        {ji_textrolspeech_2024}
\bibfield{author}{\bibinfo{person}{Shengpeng Ji}, \bibinfo{person}{Jialong Zuo}, \bibinfo{person}{Minghui Fang}, \bibinfo{person}{Ziyue Jiang}, \bibinfo{person}{Feiyang Chen}, \bibinfo{person}{Xinyu Duan}, \bibinfo{person}{Baoxing Huai}, {and} \bibinfo{person}{Zhou Zhao}.} \bibinfo{year}{2024}\natexlab{}.
\newblock \showarticletitle{{TextrolSpeech}: {A} {Text} {Style} {Control} {Speech} {Corpus} with {Codec} {Language} {Text}-to-{Speech} {Models}}. In \bibinfo{booktitle}{\emph{{ICASSP} 2024 - 2024 {IEEE} {International} {Conference} on {Acoustics}, {Speech} and {Signal} {Processing} ({ICASSP})}}. \bibinfo{pages}{10301--10305}.
\newblock
\urldef\tempurl%
\url{https://doi.org/10.1109/ICASSP48485.2024.10445879}
\showDOI{\tempurl}


\bibitem[Jin et~al\mbox{.}(2023)]%
        {jin2023holosinger}
\bibfield{author}{\bibinfo{person}{Zeyu Jin}, \bibinfo{person}{Zixuan Wang}, \bibinfo{person}{Qixin Wang}, \bibinfo{person}{Jia Jia}, \bibinfo{person}{Ye Bai}, \bibinfo{person}{Yi Zhao}, \bibinfo{person}{Hao Li}, {and} \bibinfo{person}{Xiaorui Wang}.} \bibinfo{year}{2023}\natexlab{}.
\newblock \showarticletitle{HoloSinger: Semantics and Music Driven Motion Generation with Octahedral Holographic Projection}. In \bibinfo{booktitle}{\emph{Proceedings of the 31st ACM International Conference on Multimedia}}. \bibinfo{pages}{9393--9395}.
\newblock


\bibitem[Kearns(2014)]%
        {kearns_librivox_2014}
\bibfield{author}{\bibinfo{person}{Jodi Kearns}.} \bibinfo{year}{2014}\natexlab{}.
\newblock \showarticletitle{Librivox: {Free} public domain audiobooks}.
\newblock \bibinfo{journal}{\emph{Reference Reviews}} \bibinfo{volume}{28}, \bibinfo{number}{1} (\bibinfo{year}{2014}), \bibinfo{pages}{7--8}.
\newblock
\urldef\tempurl%
\url{https://doi.org/10.1108/RR-08-2013-0197}
\showDOI{\tempurl}
\newblock
\shownote{Publisher: Emerald group publishing limited}.


\bibitem[Kim et~al\mbox{.}(2019)]%
        {kim_audiocaps_2019}
\bibfield{author}{\bibinfo{person}{Chris~Dongjoo Kim}, \bibinfo{person}{Byeongchang Kim}, \bibinfo{person}{Hyunmin Lee}, {and} \bibinfo{person}{Gunhee Kim}.} \bibinfo{year}{2019}\natexlab{}.
\newblock \showarticletitle{{AudioCaps}: {Generating} {Captions} for {Audios} in {The} {Wild}}. In \bibinfo{booktitle}{\emph{Proceedings of the 2019 {Conference} of the {North} {American} {Chapter} of the {Association} for {Computational} {Linguistics}: {Human} {Language} {Technologies}, {Volume} 1 ({Long} and {Short} {Papers})}}. \bibinfo{publisher}{Association for Computational Linguistics}, \bibinfo{address}{Minneapolis, Minnesota}, \bibinfo{pages}{119--132}.
\newblock
\urldef\tempurl%
\url{https://doi.org/10.18653/v1/N19-1011}
\showDOI{\tempurl}


\bibitem[Kim et~al\mbox{.}(2021)]%
        {kim_expressive_2021}
\bibfield{author}{\bibinfo{person}{Minchan Kim}, \bibinfo{person}{Sung~Jun Cheon}, \bibinfo{person}{Byoung~Jin Choi}, \bibinfo{person}{Jong~Jin Kim}, {and} \bibinfo{person}{Nam~Soo Kim}.} \bibinfo{year}{2021}\natexlab{}.
\newblock \showarticletitle{Expressive text-to-speech using style tag}.
\newblock \bibinfo{journal}{\emph{arXiv preprint arXiv:2104.00436}} (\bibinfo{year}{2021}).
\newblock


\bibitem[Koepke et~al\mbox{.}(2023)]%
        {koepke_audio_2023}
\bibfield{author}{\bibinfo{person}{A.~Sophia Koepke}, \bibinfo{person}{Andreea-Maria Oncescu}, \bibinfo{person}{João~F. Henriques}, \bibinfo{person}{Zeynep Akata}, {and} \bibinfo{person}{Samuel Albanie}.} \bibinfo{year}{2023}\natexlab{}.
\newblock \showarticletitle{Audio {Retrieval} {With} {Natural} {Language} {Queries}: {A} {Benchmark} {Study}}.
\newblock \bibinfo{journal}{\emph{IEEE Transactions on Multimedia}}  \bibinfo{volume}{25} (\bibinfo{year}{2023}), \bibinfo{pages}{2675--2685}.
\newblock
\urldef\tempurl%
\url{https://doi.org/10.1109/TMM.2022.3149712}
\showDOI{\tempurl}


\bibitem[Koizumi et~al\mbox{.}(2023)]%
        {koizumi_libritts-r_2023}
\bibfield{author}{\bibinfo{person}{Yuma Koizumi}, \bibinfo{person}{Heiga Zen}, \bibinfo{person}{Shigeki Karita}, \bibinfo{person}{Yifan Ding}, \bibinfo{person}{Kohei Yatabe}, \bibinfo{person}{Nobuyuki Morioka}, \bibinfo{person}{Michiel Bacchiani}, \bibinfo{person}{Yu Zhang}, \bibinfo{person}{Wei Han}, {and} \bibinfo{person}{Ankur Bapna}.} \bibinfo{year}{2023}\natexlab{}.
\newblock \showarticletitle{Libritts-r: {A} restored multi-speaker text-to-speech corpus}.
\newblock \bibinfo{journal}{\emph{arXiv preprint arXiv:2305.18802}} (\bibinfo{year}{2023}).
\newblock


\bibitem[Kong et~al\mbox{.}(2023)]%
        {kong_vits2_2023}
\bibfield{author}{\bibinfo{person}{Jungil Kong}, \bibinfo{person}{Jihoon Park}, \bibinfo{person}{Beomjeong Kim}, \bibinfo{person}{Jeongmin Kim}, \bibinfo{person}{Dohee Kong}, {and} \bibinfo{person}{Sangjin Kim}.} \bibinfo{year}{2023}\natexlab{}.
\newblock \showarticletitle{{VITS2}: {Improving} {Quality} and {Efficiency} of {Single}-{Stage} {Text}-to-{Speech} with {Adversarial} {Learning} and {Architecture} {Design}}.
\newblock \bibinfo{journal}{\emph{arXiv preprint arXiv:2307.16430}} (\bibinfo{year}{2023}).
\newblock


\bibitem[Kumar and Raj(2016)]%
        {kumar_audio_2016}
\bibfield{author}{\bibinfo{person}{Anurag Kumar} {and} \bibinfo{person}{Bhiksha Raj}.} \bibinfo{year}{2016}\natexlab{}.
\newblock \showarticletitle{Audio {Event} {Detection} using {Weakly} {Labeled} {Data}}. In \bibinfo{booktitle}{\emph{Proceedings of the 24th {ACM} {International} {Conference} on {Multimedia}}} \emph{(\bibinfo{series}{{MM} '16})}. \bibinfo{publisher}{Association for Computing Machinery}, \bibinfo{address}{New York, NY, USA}, \bibinfo{pages}{1038--1047}.
\newblock
\showISBNx{978-1-4503-3603-1}
\urldef\tempurl%
\url{https://doi.org/10.1145/2964284.2964310}
\showDOI{\tempurl}
\newblock
\shownote{event-place: Amsterdam, The Netherlands}.


\bibitem[Leng et~al\mbox{.}(2023)]%
        {leng_prompttts_2023}
\bibfield{author}{\bibinfo{person}{Yichong Leng}, \bibinfo{person}{Zhifang Guo}, \bibinfo{person}{Kai Shen}, \bibinfo{person}{Xu Tan}, \bibinfo{person}{Zeqian Ju}, \bibinfo{person}{Yanqing Liu}, \bibinfo{person}{Yufei Liu}, \bibinfo{person}{Dongchao Yang}, \bibinfo{person}{Leying Zhang}, \bibinfo{person}{Kaitao Song}, \bibinfo{person}{Lei He}, \bibinfo{person}{Xiang-Yang Li}, \bibinfo{person}{Sheng Zhao}, \bibinfo{person}{Tao Qin}, {and} \bibinfo{person}{Jiang Bian}.} \bibinfo{year}{2023}\natexlab{}.
\newblock \showarticletitle{{PromptTTS} 2: {Describing} and {Generating} {Voices} with {Text} {Prompt}}.
\newblock \bibinfo{journal}{\emph{arXiv preprint arXiv:2309.02285}} (\bibinfo{year}{2023}).
\newblock


\bibitem[Lyth and King(2024)]%
        {lyth2024natural}
\bibfield{author}{\bibinfo{person}{Dan Lyth} {and} \bibinfo{person}{Simon King}.} \bibinfo{year}{2024}\natexlab{}.
\newblock \showarticletitle{Natural language guidance of high-fidelity text-to-speech with synthetic annotations}.
\newblock \bibinfo{journal}{\emph{arXiv preprint arXiv:2402.01912}} (\bibinfo{year}{2024}).
\newblock


\bibitem[Ma et~al\mbox{.}(2023)]%
        {ma_emotion2vec_2023}
\bibfield{author}{\bibinfo{person}{Ziyang Ma}, \bibinfo{person}{Zhisheng Zheng}, \bibinfo{person}{Jiaxin Ye}, \bibinfo{person}{Jinchao Li}, \bibinfo{person}{Zhifu Gao}, \bibinfo{person}{Shiliang Zhang}, {and} \bibinfo{person}{Xie Chen}.} \bibinfo{year}{2023}\natexlab{}.
\newblock \showarticletitle{emotion2vec: {Self}-{Supervised} {Pre}-{Training} for {Speech} {Emotion} {Representation}}.
\newblock \bibinfo{journal}{\emph{arXiv preprint arXiv:2312.15185}} (\bibinfo{year}{2023}).
\newblock


\bibitem[Mei et~al\mbox{.}(2023)]%
        {mei_wavcaps_2023}
\bibfield{author}{\bibinfo{person}{Xinhao Mei}, \bibinfo{person}{Chutong Meng}, \bibinfo{person}{Haohe Liu}, \bibinfo{person}{Qiuqiang Kong}, \bibinfo{person}{Tom Ko}, \bibinfo{person}{Chengqi Zhao}, \bibinfo{person}{Mark~D Plumbley}, \bibinfo{person}{Yuexian Zou}, {and} \bibinfo{person}{Wenwu Wang}.} \bibinfo{year}{2023}\natexlab{}.
\newblock \showarticletitle{Wavcaps: {A} chatgpt-assisted weakly-labelled audio captioning dataset for audio-language multimodal research}.
\newblock \bibinfo{journal}{\emph{arXiv preprint arXiv:2303.17395}} (\bibinfo{year}{2023}).
\newblock


\bibitem[Mihalcea and Tarau(2004)]%
        {mihalcea2004textrank}
\bibfield{author}{\bibinfo{person}{Rada Mihalcea} {and} \bibinfo{person}{Paul Tarau}.} \bibinfo{year}{2004}\natexlab{}.
\newblock \showarticletitle{Textrank: Bringing order into text}. In \bibinfo{booktitle}{\emph{Proceedings of the 2004 conference on empirical methods in natural language processing}}. \bibinfo{pages}{404--411}.
\newblock


\bibitem[OpenAI(2024)]%
        {openai_gpt-4_2024}
\bibfield{author}{\bibinfo{person}{OpenAI}.} \bibinfo{year}{2024}\natexlab{}.
\newblock \showarticletitle{{GPT}-4 {Technical} {Report}}.
\newblock \bibinfo{journal}{\emph{arXiv preprint arXiv:2303.08774}} (\bibinfo{year}{2024}).
\newblock


\bibitem[Radford et~al\mbox{.}(2021)]%
        {radford_learning_2021}
\bibfield{author}{\bibinfo{person}{Alec Radford}, \bibinfo{person}{Jong~Wook Kim}, \bibinfo{person}{Chris Hallacy}, \bibinfo{person}{Aditya Ramesh}, \bibinfo{person}{Gabriel Goh}, \bibinfo{person}{Sandhini Agarwal}, \bibinfo{person}{Girish Sastry}, \bibinfo{person}{Amanda Askell}, \bibinfo{person}{Pamela Mishkin}, \bibinfo{person}{Jack Clark}, \bibinfo{person}{Gretchen Krueger}, {and} \bibinfo{person}{Ilya Sutskever}.} \bibinfo{year}{2021}\natexlab{}.
\newblock \showarticletitle{Learning {Transferable} {Visual} {Models} {From} {Natural} {Language} {Supervision}}. In \bibinfo{booktitle}{\emph{Proceedings of the 38th {International} {Conference} on {Machine} {Learning}}} \emph{(\bibinfo{series}{Proceedings of {Machine} {Learning} {Research}}, Vol.~\bibinfo{volume}{139})}. \bibinfo{publisher}{PMLR}, \bibinfo{pages}{8748--8763}.
\newblock
\urldef\tempurl%
\url{https://proceedings.mlr.press/v139/radford21a.html}
\showURL{%
\tempurl}


\bibitem[Radford et~al\mbox{.}(2022)]%
        {radford_robust_2022}
\bibfield{author}{\bibinfo{person}{Alec Radford}, \bibinfo{person}{Jong~Wook Kim}, \bibinfo{person}{Tao Xu}, \bibinfo{person}{Greg Brockman}, \bibinfo{person}{Christine McLeavey}, {and} \bibinfo{person}{Ilya Sutskever}.} \bibinfo{year}{2022}\natexlab{}.
\newblock \bibinfo{title}{Robust {Speech} {Recognition} via {Large}-{Scale} {Weak} {Supervision}}.
\newblock
\newblock
\urldef\tempurl%
\url{https://doi.org/10.48550/ARXIV.2212.04356}
\showDOI{\tempurl}


\bibitem[Ren et~al\mbox{.}(2020)]%
        {ren_fastspeech_2020}
\bibfield{author}{\bibinfo{person}{Yi Ren}, \bibinfo{person}{Chenxu Hu}, \bibinfo{person}{Xu Tan}, \bibinfo{person}{Tao Qin}, \bibinfo{person}{Sheng Zhao}, \bibinfo{person}{Zhou Zhao}, {and} \bibinfo{person}{Tie-Yan Liu}.} \bibinfo{year}{2020}\natexlab{}.
\newblock \showarticletitle{Fastspeech 2: {Fast} and high-quality end-to-end text to speech}.
\newblock \bibinfo{journal}{\emph{arXiv preprint arXiv:2006.04558}} (\bibinfo{year}{2020}).
\newblock


\bibitem[Shen et~al\mbox{.}(2018)]%
        {shen_natural_2018}
\bibfield{author}{\bibinfo{person}{Jonathan Shen}, \bibinfo{person}{Ruoming Pang}, \bibinfo{person}{Ron~J. Weiss}, \bibinfo{person}{Mike Schuster}, \bibinfo{person}{Navdeep Jaitly}, \bibinfo{person}{Zongheng Yang}, \bibinfo{person}{Zhifeng Chen}, \bibinfo{person}{Yu Zhang}, \bibinfo{person}{Yuxuan Wang}, \bibinfo{person}{Rj Skerrv-Ryan}, \bibinfo{person}{Rif~A. Saurous}, \bibinfo{person}{Yannis Agiomvrgiannakis}, {and} \bibinfo{person}{Yonghui Wu}.} \bibinfo{year}{2018}\natexlab{}.
\newblock \showarticletitle{Natural {TTS} {Synthesis} by {Conditioning} {Wavenet} on {MEL} {Spectrogram} {Predictions}}. In \bibinfo{booktitle}{\emph{2018 {IEEE} {International} {Conference} on {Acoustics}, {Speech} and {Signal} {Processing} ({ICASSP})}}. \bibinfo{pages}{4779--4783}.
\newblock
\urldef\tempurl%
\url{https://doi.org/10.1109/ICASSP.2018.8461368}
\showDOI{\tempurl}


\bibitem[Shen et~al\mbox{.}(2023)]%
        {shen_naturalspeech_2023}
\bibfield{author}{\bibinfo{person}{Kai Shen}, \bibinfo{person}{Zeqian Ju}, \bibinfo{person}{Xu Tan}, \bibinfo{person}{Yanqing Liu}, \bibinfo{person}{Yichong Leng}, \bibinfo{person}{Lei He}, \bibinfo{person}{Tao Qin}, \bibinfo{person}{Sheng Zhao}, {and} \bibinfo{person}{Jiang Bian}.} \bibinfo{year}{2023}\natexlab{}.
\newblock \showarticletitle{Naturalspeech 2: {Latent} diffusion models are natural and zero-shot speech and singing synthesizers}.
\newblock \bibinfo{journal}{\emph{arXiv preprint arXiv:2304.09116}} (\bibinfo{year}{2023}).
\newblock


\bibitem[Shi et~al\mbox{.}(2020)]%
        {shi_aishell-3_2020}
\bibfield{author}{\bibinfo{person}{Yao Shi}, \bibinfo{person}{Hui Bu}, \bibinfo{person}{Xin Xu}, \bibinfo{person}{Shaoji Zhang}, {and} \bibinfo{person}{Ming Li}.} \bibinfo{year}{2020}\natexlab{}.
\newblock \showarticletitle{Aishell-3: {A} multi-speaker mandarin tts corpus and the baselines}.
\newblock \bibinfo{journal}{\emph{arXiv preprint arXiv:2010.11567}} (\bibinfo{year}{2020}).
\newblock


\bibitem[Su(2020)]%
        {su_simbert_2020}
\bibfield{author}{\bibinfo{person}{Jianlin Su}.} \bibinfo{year}{2020}\natexlab{}.
\newblock \bibinfo{booktitle}{\emph{{SimBERT}: {Integrating} {Retrieval} and {Generation} into {BERT}}}.
\newblock \bibinfo{type}{{T}echnical {R}eport}.
\newblock
\urldef\tempurl%
\url{https://github.com/ZhuiyiTechnology/simbert}
\showURL{%
\tempurl}


\bibitem[Sun et~al\mbox{.}(2024)]%
        {sun2024video}
\bibfield{author}{\bibinfo{person}{Guangzhi Sun}, \bibinfo{person}{Wenyi Yu}, \bibinfo{person}{Changli Tang}, \bibinfo{person}{Xianzhao Chen}, \bibinfo{person}{Tian Tan}, \bibinfo{person}{Wei Li}, \bibinfo{person}{Lu Lu}, \bibinfo{person}{Zejun Ma}, \bibinfo{person}{Yuxuan Wang}, {and} \bibinfo{person}{Chao Zhang}.} \bibinfo{year}{2024}\natexlab{}.
\newblock \showarticletitle{video-SALMONN: Speech-Enhanced Audio-Visual Large Language Models}.
\newblock \bibinfo{journal}{\emph{arXiv preprint arXiv:2406.15704}} (\bibinfo{year}{2024}).
\newblock


\bibitem[Tang et~al\mbox{.}(2023)]%
        {tang_salmonn_2023}
\bibfield{author}{\bibinfo{person}{Changli Tang}, \bibinfo{person}{Wenyi Yu}, \bibinfo{person}{Guangzhi Sun}, \bibinfo{person}{Xianzhao Chen}, \bibinfo{person}{Tian Tan}, \bibinfo{person}{Wei Li}, \bibinfo{person}{Lu Lu}, \bibinfo{person}{Zejun Ma}, {and} \bibinfo{person}{Chao Zhang}.} \bibinfo{year}{2023}\natexlab{}.
\newblock \showarticletitle{Salmonn: {Towards} generic hearing abilities for large language models}.
\newblock \bibinfo{journal}{\emph{arXiv preprint arXiv:2310.13289}} (\bibinfo{year}{2023}).
\newblock


\bibitem[Touvron et~al\mbox{.}(2023)]%
        {touvron_llama_2023}
\bibfield{author}{\bibinfo{person}{Hugo Touvron}, \bibinfo{person}{Louis Martin}, \bibinfo{person}{Kevin Stone}, \bibinfo{person}{Peter Albert}, \bibinfo{person}{Amjad Almahairi}, \bibinfo{person}{Yasmine Babaei}, \bibinfo{person}{Nikolay Bashlykov}, \bibinfo{person}{Soumya Batra}, \bibinfo{person}{Prajjwal Bhargava}, \bibinfo{person}{Shruti Bhosale}, {and} \bibinfo{person}{{others}}.} \bibinfo{year}{2023}\natexlab{}.
\newblock \showarticletitle{Llama 2: {Open} foundation and fine-tuned chat models}.
\newblock \bibinfo{journal}{\emph{arXiv preprint arXiv:2307.09288}} (\bibinfo{year}{2023}).
\newblock


\bibitem[Vlasenko et~al\mbox{.}(2007)]%
        {vlasenko_combining_2007}
\bibfield{author}{\bibinfo{person}{Bogdan Vlasenko}, \bibinfo{person}{Björn Schuller}, \bibinfo{person}{Andreas Wendemuth}, {and} \bibinfo{person}{Gerhard Rigoll}.} \bibinfo{year}{2007}\natexlab{}.
\newblock \showarticletitle{Combining frame and turn-level information for robust recognition of emotions within speech}. In \bibinfo{booktitle}{\emph{Proc. {INTERSPEECH} 2007, {Antwerp}, {Belgium}}}.
\newblock
\urldef\tempurl%
\url{https://doi.org/10.21437/Interspeech.2007-611}
\showDOI{\tempurl}


\bibitem[Vrehuuvrek and Sojka(2010)]%
        {vrehuuvrek2010software}
\bibfield{author}{\bibinfo{person}{Radim Vrehuuvrek} {and} \bibinfo{person}{Petr Sojka}.} \bibinfo{year}{2010}\natexlab{}.
\newblock \showarticletitle{Software framework for topic modelling with large corpora}.
\newblock  (\bibinfo{year}{2010}).
\newblock


\bibitem[Vyas et~al\mbox{.}(2023)]%
        {vyas_audiobox_2023}
\bibfield{author}{\bibinfo{person}{Apoorv Vyas}, \bibinfo{person}{Bowen Shi}, \bibinfo{person}{Matthew Le}, \bibinfo{person}{Andros Tjandra}, \bibinfo{person}{Yi-Chiao Wu}, \bibinfo{person}{Baishan Guo}, \bibinfo{person}{Jiemin Zhang}, \bibinfo{person}{Xinyue Zhang}, \bibinfo{person}{Robert Adkins}, \bibinfo{person}{William Ngan}, {and} \bibinfo{person}{{others}}.} \bibinfo{year}{2023}\natexlab{}.
\newblock \showarticletitle{Audiobox: {Unified} audio generation with natural language prompts}.
\newblock \bibinfo{journal}{\emph{arXiv preprint arXiv:2312.15821}} (\bibinfo{year}{2023}).
\newblock


\bibitem[Wang et~al\mbox{.}(2023)]%
        {wang_neural_2023}
\bibfield{author}{\bibinfo{person}{Chengyi Wang}, \bibinfo{person}{Sanyuan Chen}, \bibinfo{person}{Yu Wu}, \bibinfo{person}{Ziqiang Zhang}, \bibinfo{person}{Long Zhou}, \bibinfo{person}{Shujie Liu}, \bibinfo{person}{Zhuo Chen}, \bibinfo{person}{Yanqing Liu}, \bibinfo{person}{Huaming Wang}, \bibinfo{person}{Jinyu Li}, {and} \bibinfo{person}{{others}}.} \bibinfo{year}{2023}\natexlab{}.
\newblock \showarticletitle{Neural codec language models are zero-shot text to speech synthesizers.(2023)}.
\newblock \bibinfo{journal}{\emph{arXiv preprint arXiv:2301.02111}} (\bibinfo{year}{2023}).
\newblock


\bibitem[Wang et~al\mbox{.}(2020)]%
        {wang_mead_2020}
\bibfield{author}{\bibinfo{person}{Kaisiyuan Wang}, \bibinfo{person}{Qianyi Wu}, \bibinfo{person}{Linsen Song}, \bibinfo{person}{Zhuoqian Yang}, \bibinfo{person}{Wayne Wu}, \bibinfo{person}{Chen Qian}, \bibinfo{person}{Ran He}, \bibinfo{person}{Yu Qiao}, {and} \bibinfo{person}{Chen~Change Loy}.} \bibinfo{year}{2020}\natexlab{}.
\newblock \showarticletitle{{MEAD}: {A} {Large}-{Scale} {Audio}-{Visual} {Dataset} for {Emotional} {Talking}-{Face} {Generation}}. In \bibinfo{booktitle}{\emph{Computer {Vision} – {ECCV} 2020}}. \bibinfo{publisher}{Springer International Publishing}, \bibinfo{address}{Cham}, \bibinfo{pages}{700--717}.
\newblock
\showISBNx{978-3-030-58589-1}


\bibitem[Wang et~al\mbox{.}(2024)]%
        {wang2024dancecamera3d}
\bibfield{author}{\bibinfo{person}{Zixuan Wang}, \bibinfo{person}{Jia Jia}, \bibinfo{person}{Shikun Sun}, \bibinfo{person}{Haozhe Wu}, \bibinfo{person}{Rong Han}, \bibinfo{person}{Zhenyu Li}, \bibinfo{person}{Di Tang}, \bibinfo{person}{Jiaqing Zhou}, {and} \bibinfo{person}{Jiebo Luo}.} \bibinfo{year}{2024}\natexlab{}.
\newblock \showarticletitle{DanceCamera3D: 3D Camera Movement Synthesis with Music and Dance}. In \bibinfo{booktitle}{\emph{Proceedings of the IEEE/CVF Conference on Computer Vision and Pattern Recognition}}. \bibinfo{pages}{7892--7901}.
\newblock


\bibitem[Wu et~al\mbox{.}(2023)]%
        {wu_large-scale_2023}
\bibfield{author}{\bibinfo{person}{Yusong Wu}, \bibinfo{person}{Ke Chen}, \bibinfo{person}{Tianyu Zhang}, \bibinfo{person}{Yuchen Hui}, \bibinfo{person}{Taylor Berg-Kirkpatrick}, {and} \bibinfo{person}{Shlomo Dubnov}.} \bibinfo{year}{2023}\natexlab{}.
\newblock \showarticletitle{Large-{Scale} {Contrastive} {Language}-{Audio} {Pretraining} with {Feature} {Fusion} and {Keyword}-to-{Caption} {Augmentation}}. In \bibinfo{booktitle}{\emph{{ICASSP} 2023 - 2023 {IEEE} {International} {Conference} on {Acoustics}, {Speech} and {Signal} {Processing} ({ICASSP})}}. \bibinfo{pages}{1--5}.
\newblock
\urldef\tempurl%
\url{https://doi.org/10.1109/ICASSP49357.2023.10095969}
\showDOI{\tempurl}


\bibitem[Xu et~al\mbox{.}(2023)]%
        {xu_multimodal_2023}
\bibfield{author}{\bibinfo{person}{Peng Xu}, \bibinfo{person}{Xiatian Zhu}, {and} \bibinfo{person}{David~A. Clifton}.} \bibinfo{year}{2023}\natexlab{}.
\newblock \showarticletitle{Multimodal {Learning} {With} {Transformers}: {A} {Survey}}.
\newblock \bibinfo{journal}{\emph{IEEE Transactions on Pattern Analysis and Machine Intelligence}} \bibinfo{volume}{45}, \bibinfo{number}{10} (\bibinfo{year}{2023}), \bibinfo{pages}{12113--12132}.
\newblock
\urldef\tempurl%
\url{https://doi.org/10.1109/TPAMI.2023.3275156}
\showDOI{\tempurl}


\bibitem[Xu et~al\mbox{.}(2024)]%
        {xu_secap_2024}
\bibfield{author}{\bibinfo{person}{Yaoxun Xu}, \bibinfo{person}{Hangting Chen}, \bibinfo{person}{Jianwei Yu}, \bibinfo{person}{Qiaochu Huang}, \bibinfo{person}{Zhiyong Wu}, \bibinfo{person}{Shi-Xiong Zhang}, \bibinfo{person}{Guangzhi Li}, \bibinfo{person}{Yi Luo}, {and} \bibinfo{person}{Rongzhi Gu}.} \bibinfo{year}{2024}\natexlab{}.
\newblock \showarticletitle{Secap: {Speech} emotion captioning with large language model}. In \bibinfo{booktitle}{\emph{Proceedings of the {AAAI} {Conference} on {Artificial} {Intelligence}}}, Vol.~\bibinfo{volume}{38}. \bibinfo{pages}{19323--19331}.
\newblock
\newblock
\shownote{Issue: 17}.


\bibitem[Yang and et~al(2023)]%
        {yang_baichuan_2023}
\bibfield{author}{\bibinfo{person}{Aiyuan Yang} {and} \bibinfo{person}{et al}.} \bibinfo{year}{2023}\natexlab{}.
\newblock \showarticletitle{Baichuan 2: {Open} {Large}-scale {Language} {Models}}.
\newblock \bibinfo{journal}{\emph{arXiv preprint arXiv:2309.10305}} (\bibinfo{year}{2023}).
\newblock


\bibitem[Yang et~al\mbox{.}(2023)]%
        {yang_instructtts_2023}
\bibfield{author}{\bibinfo{person}{Dongchao Yang}, \bibinfo{person}{Songxiang Liu}, \bibinfo{person}{Rongjie Huang}, \bibinfo{person}{Chao Weng}, {and} \bibinfo{person}{Helen Meng}.} \bibinfo{year}{2023}\natexlab{}.
\newblock \showarticletitle{Instructtts: {Modelling} expressive {TTS} in discrete latent space with natural language style prompt}.
\newblock \bibinfo{journal}{\emph{arXiv preprint arXiv:2301.13662}} (\bibinfo{year}{2023}).
\newblock


\bibitem[Zen et~al\mbox{.}(2019)]%
        {zen_libritts_2019}
\bibfield{author}{\bibinfo{person}{Heiga Zen}, \bibinfo{person}{Viet Dang}, \bibinfo{person}{Rob Clark}, \bibinfo{person}{Yu Zhang}, \bibinfo{person}{Ron~J Weiss}, \bibinfo{person}{Ye Jia}, \bibinfo{person}{Zhifeng Chen}, {and} \bibinfo{person}{Yonghui Wu}.} \bibinfo{year}{2019}\natexlab{}.
\newblock \showarticletitle{Libritts: {A} corpus derived from librispeech for text-to-speech}.
\newblock \bibinfo{journal}{\emph{arXiv preprint arXiv:1904.02882}} (\bibinfo{year}{2019}).
\newblock


\bibitem[Zhou et~al\mbox{.}(2017)]%
        {zhou_machine_2017}
\bibfield{author}{\bibinfo{person}{Lina Zhou}, \bibinfo{person}{Shimei Pan}, \bibinfo{person}{Jianwu Wang}, {and} \bibinfo{person}{Athanasios~V. Vasilakos}.} \bibinfo{year}{2017}\natexlab{}.
\newblock \showarticletitle{Machine learning on big data: {Opportunities} and challenges}.
\newblock \bibinfo{journal}{\emph{Neurocomputing}}  \bibinfo{volume}{237} (\bibinfo{year}{2017}), \bibinfo{pages}{350--361}.
\newblock
\showISSN{0925-2312}
\urldef\tempurl%
\url{https://doi.org/10.1016/j.neucom.2017.01.026}
\showDOI{\tempurl}


\end{thebibliography}

\newpage

\appendix











\section{Details in Speech Annotation}

\subsection{Example of Metadata}

In Section 3.2 of the paper, we conduct data preprocessing to establish standard metadata before the whole annotation framework and further utilize the raw metadata to extract prior information of the audio clip such as the topic. An example of preprocessed metadata is shown as follows.

\begin{lstlisting}[language=Python]
{
  'path': './giga_00000616/POD0000013293_S0000160.wav',  
  'subdatasets': 'POD',  
  'sampling_rate': 16000,  
  'title': 'Law_Report_A_decade_since_911, _54_new_anti-terror_laws',  
  'url': 'https://abcmedia.akamaized.net/rn/podcast/2011/09/lrt_20110906_0845.mp3',  
  'sid': 'POD0000013293_S0000160',  
  'speaker': 'N/A',  
  'begin_time': 0,  
  'end_time': 10.220000000000027,  
  'text_raw': "And that can actually fuel the possibilities of extremism and recruitment by terrorists . and that's why the laws need to be balanced out with other programs which enhance social cohesion , ", 
  'category': 'News and Politics'
}
\end{lstlisting}

\subsection{Label Categories}

As described in Section 3.3, our proposed annotation system characterized speech in terms of various style properties including pitch, energy, speed, age, gender, emotion description, and word emphasis. The subset labels of each attribute are listed below.

\noindent\textbf{Gender}: Male,  Female

\noindent\textbf{Age}: Child, Teenager, Youth adult, Middle-aged, Elderly

\noindent\textbf{Pitch}: low, normal, high

\noindent\textbf{Speed}: slow, normal, fast

\noindent\textbf{Volume}: low, normal, high

\noindent\textbf{Emotion} (English): Fearful, Happy, Disgusted, Sad, Surprised, Angry, Neutral

\section{SpeechCraft Data Sources}
As illustrated in Section 4.1, we implemented the annotation system across large-scale bilingual speech datasets to conduct speech descriptions, resulting in the SpeechCraft dataset.
The detailed information of the four open-source speech datasets is introduced as follows.

\noindent\textbf{AISHELL-3}~\cite{shi_aishell-3_2020} is a high-fidelity Mandarin speech corpus containing roughly 85 hours of recordings spoken by 218 speakers and a total of 88, 035 utterances. 

\noindent\textbf{Zhvoice\footnote{https://github.com/fighting41love/zhvoice}} corpus consists of eight open-source subdatasets,  processed through noise reduction and quality enhancement, with approximately 3200 speakers, and 900 hours of audio, totally 1,032,940 utterances. 

\noindent\textbf{LibriTTS-R}~\cite{koizumi_libritts-r_2023} is a restored English TTS Corpus with 585 hours of speech from 2, 456 speakers. It is derived by applying speech restoration to the LibriTTS~\cite{zen_libritts_2019}corpus with sound quality improved.

\noindent\textbf{GigaSpeech}~\cite{chen_gigaspeech_2021} is an evolving,  multi-domain ASR Corpus collected from three different sources of data,  including audiobook,  podcast,  and YouTube videos. It provides a wide range of choices on the dataset scale. The GigaSpeech-m is an officially recommended subset as a 1000-hour dataset for research experiments.

\section{Details in Emphasis Studies}

\subsection{Emphasis Regeneration}
As introduced in Section 4.2, we employed FastSpeech 2~\cite{ren_fastspeech_2020} as the backbone to regenerate the AISHELL-3 and LibriTTS-R datasets with keyword emphasized in each piece.

\subsubsection{FastSpeech~2 Backbone Model. }

FastSpeech 2 achieved modulation of phoneme-level characteristics with the key component of Variance Adaptor. 
It consists of three primary Predictors for energy, pitch and duration respectively. Adjusting the output of Predictors by scale, we can obtain loud volume, high pitch, and elongated sounds on the designed phoneme. The different combination of scaling factors determines various acoustic effects.

\subsubsection{Keyword Extraction. }

We used the TextRank~\cite{mihalcea2004textrank} algorithm for Chinese content and Gensim~\cite{vrehuuvrek2010software} for English to conduct keyword extraction.
Words such as particles and proper nouns are overlooked as they are seldom stressed in conversational speech.

\subsection{Emphasis Detection}
We introduced a word-level emphasis detection model in Section 3.3. 
The detection model works by conducting forced alignment for each waveform to get the separate audio slice for a minimum word unit.
As the emphasis is a relative concept that becomes significant only when compared to the surrounding words, we concatenate features of predecessor and successor to form the final representation for each audio unit in the neural network training.

\subsection{Emphasis Evaluation on Real-Life Dataset}
In Section 4.3, we demonstrated the accuracy of emphasis detection on the testset of the regenerated AISHELL-3-stressed and LibriTTS-R-stressed data.
To further evaluate the effectiveness of our detection approach in modeling real-life stress patterns, we utilized an internal dataset with human annotation on word emphasis over 10, 000 audio utterances read by professional voice actors to test the the model's generalization ability on real-world data.
The emphasis detection models achieve 66.90\% on word-level accuracy and 41.63\% on sentence-level accuracy on the human-annotated dataset, which showcase the model's promising ability to generalize on natural word emphasis.
However, the limited accuracy may stem from the inherent complexity of real-world emphasizing effects, which are more nuanced and varied than the mixed feature adjustments in our data construction.

\section{Model Configuration in TTS Experiments}
We adopted the Salle~\cite{ji_textrolspeech_2024} model to conduct Expressive Speech Synthesis in Section 5.1, which facilitate speech synthesis task by codec language modelling with RVQ~\cite{défossez2022highfidelityneuralaudio}.
RVQ is a method of high fidelity neural audio compression which features multi-layer discrete quantizers that can be reconstructed to high-quality waveforms by the pre-trained neural audio codec model. 

Salle employed a hybrid approach combining an autoregressive style conditional codec model (AR) and a non-autoregressive TTS codec model (NAR). 
The AR model is employed to generate the first layer of RVQ, which encapsulates the fundamental speaking information. Conversely, the NAR model is reserved for the subsequent layers of quantizers that capture fine acoustic details. 
Text style prompt embedding is concatenated ahead of the phoneme text embedding, serving as the condition. The strategic embedding integration significantly enhances the expressive capacity and accuracy of the speech synthesis.

\section{LLM Prompt}

The text prompt given to the LLM for label rewriting is as followed.

\textit{Given the pitch, volume, age, gender, tone, and text, use sentiment analysis techniques to describe in natural language what age, what gender of a person, with what kind of emotion and tone, using what kind of pitch and volume, spoke the words in the text.}

\textit{Note: You must vividly describe the sentence's intonation, pitch, tone, and emotion. All outputs must strictly avoid identical wording and sentence structure. There is no need to describe body language or psychological state, no need to emphasize specific words, and do not repeat the input content.}

\textit{Refer to the format of the following four cases:}

\textit{\textbf{Example Input - Example Output}}

\textit{Now try to process the following sentences, directly output the converted sentences according to the examples without missing any labels.}




\end{document}